\documentclass[preprint,authoryear, 5p]{elsarticle}
\usepackage{amsmath}
\usepackage{units}
\usepackage{color}
\usepackage[para]{threeparttable}
\usepackage[switch,columnwise]{lineno}
\journal{Planetary and Space Science}
\begin{document}
\begin{frontmatter}

\title{3D climate modeling of Earth-like extrasolar planets orbiting different types of host stars}
\author[dlr,zaa]{M.~Godolt\corref{cor1}}
\cortext[cor1]{Corresponding author\\ Email: mareike.godolt@dlr.de\\Phone: +49-(0)30-67055202}
\author[dlr]{J.~L.~Grenfell}
\author[fu]{A.~Hamann-Reinus}
\author[zaa,be] {D.~Kitzmann}
\author[fu]{M.~Kunze}
\author[fu]{U.~Langematz}
\author[dlr,bo]{P.~von Paris}
\author[zaa]{A.~B.~C.~Patzer}
\author[dlr,zaa]{H.~Rauer}
\author[dlr]{B.~Stracke}
\address[dlr]{Extrasolare Planeten und Atmosph\"aren, Institut f\"ur Planetenforschung, Deutsches Zentrum f\"ur Luft- und Raumfahrt, Rutherfordstr.2, 12489 Berlin,Germany}
\address[zaa]{Zentrum f\"ur Astronomie und Astrophysik, Technische Universit\"at Berlin, Hardenbergstr. 36, 10623 Berlin, Germany}
\address[fu]{Institut f\"ur Meteorologie, Freie Universit\"at Berlin, Carl-Heinrich-Becker-Weg 6-10, 12165 Berlin, Germany}
\address[be]{now at Center for Space and Habitability (CSH), Sidlerstrasse 5, 3012 Bern, Switzerland}
\address[bo]{now at Universit\'e de Bordeaux, LAB, UMR 5804, F-33270, Floirac, France, and CNRS, LAB, UMR 5804, F-33270, Floirac, France }
\begin{abstract}

%%%Motivation
The potential habitability of a terrestrial planet is usually defined by the possible existence of liquid water on its surface, since life as we know it needs liquid water at least during a part of its life cycle. The potential presence of liquid water on a planetary surface depends on many factors such as, most importantly, surface temperatures. The properties of the planetary atmosphere and its interaction with the radiative energy provided by the planet's host star are thereby of decisive importance.\\
%%Method
In this study we investigate the influence of different main-sequence stars (F, G, and K-type stars) upon the climate of Earth-like extrasolar planets and their potential habitability by applying a state-of-the-art three-dimensional (3D) Earth climate model accounting for local and dynamical processes. 
The calculations have been performed for planets with Earth-like atmospheres at orbital distances (and corresponding orbital periods) where the total amount of energy received from the various host stars equals the solar constant. In contrast to previous 3D modeling studies, we include the effect of ozone radiative heating upon the vertical temperature structure of the atmospheres. The global orbital mean results obtained have been compared to those of a one-dimensional (1D) radiative convective climate model to investigate the approximation of global mean 3D results by those of 1D models.\\
%%%Results
The different stellar spectral energy distributions lead to different surface temperatures and due to ozone heating to very different vertical temperature structures.
As previous 1D studies we find higher surface temperatures for the Earth-like planet around the K-type star, and lower temperatures for the planet around the F-type star compared to an Earth-like planet around the Sun. However, this effect is more pronounced in the 3D model results than in the 1D model because the 3D model accounts for feedback processes such as the ice-albedo and the water vapor feedback. 
Whether the 1D model may approximate the global mean of the 3D model results strongly depends on the choice of the relative humidity profile in the 1D model, which is used to determine the water vapor profile. Hence, possible changes in the hydrological cycle need to be accounted for when estimating the potential habitability of an extrasolar planet. 

\end{abstract}


\end{frontmatter}

\section{Introduction}
\noindent
As instrumental sensitivity increases the number of potentially rocky extrasolar planets detected has steadi\-ly increased since the 1990s. For these kind of objects the question of habitability, i.e.~their potential to have liquid water on the planetary surface, is of special  interest. Whether or not liquid water is in principle possible on the planetary surface depends, especially on surface temperatures and pressures, hence planetary climate. 
Therefore, from the atmospheric modeling point of view, a major objective is to identify and understand the key processes determining the planetary climate.\\
Knowledge of the nature of these small planets is rather limited, especially for those within the habitable zone. The classical habitable zone is given by the orbital distances at which liquid water is possible on the surface of an Earth-like, rocky extrasolar planet assuming e.g.~certain atmospheric compositions and masses. The boundaries of the habitable zone have been studied with 1D models, e.g.~\cite{Hart1979}, \cite{Kasting1993hz}, \cite{Forget1998HZ}, \cite{vonParis2013co2}, Pierrehumbert and Gaidos (2011)\nocite{Pierrehumbert2011H2}, \cite{Kopparapu2013}, and recently also with 3D models, e.g.~\cite{Abe2011}, Wolf and Toon (2014)\nocite{Wolf2014}, Leconte et al.~(2013b)\nocite{Leconte2013}, Yang et al.~(2014)\nocite{Yang2014}. For nearby transiting planets a mean mass density may be determined via independent measurements of planetary radius, via the transit method, and planetary mass, via the radial velocity method, as e.g. for CoRoT-7b \citep{Leger2009}, Kepler-10b (Batalha et al., 2011, Dumusque et al., 2014\nocite{Kepler10b2011,Kepler10b2014}) or GJ1214b (Charbonneau et al., 2009)\nocite{Charbonneau2009}. These kinds of independent measurements are unfortunately not yet available for potentially rocky planets within the habitable zone. However, for most of the planets and planetary candidates the orbital distance and the type of central star have been determined. This allows the estimation whether these planets lie within the so-called habitable zone. 
The influence of the stellar type upon planetary atmospheres of terrestrial planets has been studied extensively with 1D models, as e.g.~in \cite{Kasting1993hz}, \cite{Selsis2000}, \cite{Segura2003,Segura2005}, \cite{Grenfell2007}, Kitzmann et al.~(2010)\nocite{Kitzmann2010}, Rauer et al.~(2011)\nocite{Rauer2011}, Kopparapu et al.~(2013)\nocite{Kopparapu2013}, \cite{Hedelt2013}, Rug\-hei\-mer et al.~(2013)\nocite{Rugheimer2013}, showing that not only the total amount of stellar energy has a large impact on a planetary atmosphere but also its spectral energy distribution (SED). The strongly wavelength dependent absorption and scattering of stellar light by planetary atmospheres influences the surface temperature, the vertical temperature structure and also the atmospheric chemistry, which then may lead to different spectral appearances. 
3D modeling studies of terrestrial extrasolar planets have helped to gain insight into key processes and boundary conditions important for the climate of rocky extrasolar planets. These planets may be very different from Earth due to different rotation rates (Joshi et al., 1997, Joshi 2003, Yang et al., 2013, 2014\nocite{Joshi1997,Joshi2003, Yang2013,Yang2014}), different obliquities (Williams and Pollard, 2003)\nocite{Williams2003}, eccentricities (Williams and Pollard, 2002)\nocite{Williams2002}, and water reservoirs (Abe et al., 2011, Leconte et al., 2013b\nocite{Abe2011,Leconte2013}). These 3D modeling studies all indicate that especially the hydrological cycle may have a large impact on planetary habitability.\\
SEDs different from the Sun have been included in 3D atmosphere studies, such as \cite{Wordsworth2011}, Leconte et al.~(2013b)\nocite{Leconte2013}, \cite{Shields2013, Shields2014}, \cite{Yang2013,Yang2014}. The influence of the different SEDs has however only been analyzed in detail by \cite{Shields2013,Shields2014}, who investigate the influence of the stellar SED upon the ice-albedo effect. So far all 3D modeling studies including different stellar SEDs omit ozone in their atmospheres. Ozone (O$_3$) likely has a large impact on the vertical temperature structure of such atmospheres and thereby on the dynamical processes. In the search for O$_3$ as a biosignature, its effect on the temperature needs to be taken into account. Photochemistry studies by \cite{Selsis2000}, \cite{Segura2003} and Grenfell et al.~(2007)\nocite{Grenfell2007} showed that an O$_3$ layer may be expected for planets around K- and F-type stars 
with Earth-like atmospheres.\\
The impact of the SED upon the temperature structures of Earth-like atmospheres and the consequences for planetary climate have not yet been studied with a 3D climate model as previous studies neglect the influence of O$_3$ upon the temperature structures.  It will be studied in detail here.\\ 

This paper investigates the influence of the spectral stellar energy input upon Earth-like planetary atmospheres utilizing a state-of-the-art 3D General Circulation Model (GCM) of Earth. The response in temperatures, surface conditions and the hydrological cycle are analyzed and the global mean results are compared to those of a 1D cloud-free radiative-convective climate model.\\
Section \ref{sec:comp} gives details about the models used, followed by a description of the scenarios studied in section \ref{sec:scenarios}. In the results section (\ref{sec:results}) we will show the influence of the SEDs upon temperatures (\ref{sec:temperatures}), the hydrological cycle (\ref{sec:hydrocycle}), and the surface conditions (\ref{sec:surf}). The sensitivity of the 3D model results to selected model parameters is discussed in section \ref{sec:sensitivity}. In section \ref{sec:1d3d} the 3D model results are compared to those of a 1D climate model. The paper closes with a summary and conclusion (sec.~\ref{sec:sum}).

\section{Computational details}
\label{sec:comp}
\noindent
To analyze the impact of different stellar spectra upon the climate of Earth-like extrasolar planets and the importance of processes such as the water vapor and albedo feedback or atmospheric dynamics we use a 3D Earth climate model. The results are analyzed and the global orbital means are compared to those of a 1D radiative-convective climate model. Details of the atmospheric models and the modeling scenarios are described in the following text.

\subsection{3D atmospheric model for Earth-like planets}
\label{3Dmodel}
\noindent
For the 3D atmospheric model calculations of Earth-like exoplanets we use the EMAC (ECHAM/\-MESSy Atmospheric Chemistry) model (J\"ockel et al., 2006)\nocite{Joeckel2006}, which has been developed for detailed investigations of the Earth's climate. 
It uses the first version of the Modular Earth Submodel System (MESSy1) to link multi-institutional computer codes. The core atmospheric model is the 5th generation European Centre Hamburg GCM (ECHAM5,  Roeckner et al., 2006\nocite{Roeckner2006}).
We apply EMAC (ECHAM5 version 5.3.01, MESSy version 1.8) in T42\-L39-resolution, i.e.~with a spherical truncation of T42 (corresponding to a quadratic Gaussian grid of approx.~2.8$^{\circ}$ by 2.8$^{\circ}$ in latitude and longitude) and 39 hybrid pressure levels from the planetary surface up to \unit[0.01]{hPa}, which corresponds to about \unit[80]{km} for the Earth. Near the surface the grid is terrain following whereas constant pressure levels are used in the upper atmosphere.
The model setup features the key atmospheric processes determining the planetary climate, i.e.~radiative transfer, convection, the hydrological cycle including cloud processes. Atmospheric chemistry is neglected in this study, instead a fixed Earth-like atmosphere is assumed (see sec.~\ref{sec:scenarios}).\\

In EMAC the radiative transfer in the shortwave and the longwave regimes are treated separately.  In the shortwave regime the radiative transfer of the stellar radiation is calculated while in the longwave regime the transfer of the thermal radiation originating at the planetary surface and within the planetary atmosphere is treated. In the shortwave regime EMAC offers a high resolution scheme, FUBRAD (Freie Universit\"at
Berlin high-resolution RADiation scheme), originally developed for solar variability studies (Nissen et al., 2007\nocite{Nissen2007}), which operates at pressures  lower than \unit[70]{hPa}, i.e.~in the stratosphere and mesosphere.
In FUBRAD the absorption of stellar radiation by molecular oxygen (O$_2$) and O$_3$ is calculated in 49 bands ranging from \unit[121.4]{nm} to \unit[682.5]{nm}.
At higher pressures, i.e.~at lower heights, the standard radiation scheme RAD\-4\-ALL (Fou\-quart and Bonnel, 1980)\nocite{Fouquart1980} is applied, which treats the radiative transfer in the UV and visible in one band ranging from \unit[250]{nm} to \unit[690]{nm}. The shortwave radiative transfer at near-infrared (NIR) wavelengths is calculated in three bands ranging from \unit[690]{nm} to \unit[4]{$\mu$m} over the entire vertical domain.
Absorption by water (H$_2$O), carbon dioxide (CO$_2$), O$_3$, methane (CH$_4$), nitrous oxide (N$_2$O), and carbon monoxide (CO), as well as Rayleigh scattering by air, scattering by aerosols, liquid and icy cloud particles are considered. The scheme  uses the $\delta$-Eddington approximation.\\
For the calculation of the thermal radiation  in the longwave regime the Rapid Radiative Transfer Model (RRTM, Mlawer et al., 1997\nocite{RRTM}) is used, which takes into account the thermal emission of the atmosphere and planetary surface, the absorption by radiative gases (H$_2$O, CO$_2$, O$_3$, CH$_4$, N$_2$O, and chlorofluorocarbons (CFCs)), aerosols and cloud particles in 16 spectral bands (from \unit[3.08]{$\mu$m} to \unit[1000]{$\mu$m}) using the correlated-k approach.\\ 
Convective transport of dry static energy, momentum, and moisture is calculated using the ECMWF (European Centre of Me\-dium-range Wea\-ther Fore\-casts) mass-flux convection scheme \citep{Bechtold2004}. The stratiform cloud coverage is calculated based on relative humidity following \cite{Sundqvist1978}. Cloud micro physics are parametrized following Lohmann and Roeckner (1996)\nocite{Lohmann1996}. The horizontal diffusion tendency is formulated via a hyper-Laplacian based on the method by \cite{Laursen1989}, using constant diffusion coefficients, which depend on the horizontal resolution. The vertical diffusion at the surface is obtained from a bulk transfer relation. Above the surface layer, eddy diffusion is assumed using diffusion coefficients for moisture and heat, which are parametrized in terms of turbulent kinetic energy and mixing lengths \citep{Brinkop1995}. \\
At the lower boundary the atmospheric model is coupled to a mixed layer ocean model (Roeckner et al.~1995)\nocite{Roeckner1995}. It calculates the sea surface temperatures, sea ice coverage, and sea ice thickness for a mixed layer of \unit[50]{m} from the net surface heat budget and a flux correction, the so-called q-flux. It accounts for the missing horizontal and vertical heat transport in the ocean and between the ocean and the atmosphere. The flux correction has been derived from a reference scenario, see section \ref{sec:scenarios}.\\
Note that complex Earth climate models such as the one used here, rely on many complex parameterizations and parameters. Some of these parameters are used to adjust the model in such a way that it properly reproduces present day and past climate states of the Earth. Especially parameters concerning the cloud properties are often adjusted such as the asymmetry parameter of the water ice crystals or the number density of the condensation nuclei.  
This is one of the major reasons for differences in modeling results of Earth's future climate, especially on seasonal and regional scales \citep[see e.g.~][]{flato2013}. Furthermore, the vertical and horizontal resolutions could also have an influence affecting e.g.~cloud covers and precipitation patterns, as e.g.~discussed for ECHAM5 by \cite{ECHAM5-2} and for the Community Atmosphere Model 3 in \cite{Williamson2008}. Also the choice of time steps and time scales may influence the model results as shown by \cite{Williamson2013}. We usually apply a time step of \unit[300]{s} but need to reduce it occasionally during spin-up to ensure small enough temperature tendencies. 

\subsection{1D radiative-convective column model}
\label{sec:1Dmodel}
\noindent
For comparison we compute vertical global mean atmospheric temperature and water profiles with a 1D cloud-free radiative-convective column model, which ranges from the surface up to a height with a pressure of \unit[6.6$\cdot$10$^{-2}$]{hPa}. The temperature profile is calculated from energy transport by radiative transfer and convective adjustment. The radiative transfer in the shortwave regime (from 237 nm to \unit[4.5]{$\mu$m}) is solved in 38 spectral bands, using a $\delta$-Eddington approximation \citep{Toon1989} and correlated-k exponential sums. In the longwave regime either RRTM is used, as in the 3D model calculations, or  MRAC (Modified RRTM for Application in CO$_2$-dominated atmospheres, \cite{vonParis2010}) for the comparison of different relative humidity profiles, which includes  only H$_2$O and CO$_2$ in 25 bands ranging from 1 to \unit[500]{$\mu$m}.  
Whenever the lapse rate calculated via radiative equilibrium exceeds the adiabatic lapse rate, convective adjustment is performed to dry or moist adiabatic conditions. The water vapor profile is calculated from the temperature profile and a relative humidity (RH) parametrization by Manabe and Wetherald (1967)\nocite{MW1967}. For the  comparison of different relative humidities (see section \ref{sec:1d3d}), additionally,  a fully saturated atmosphere is assumed (RH=100\%). A detailed description of the atmospheric model is given in \cite{Rauer2011} and \cite{vonParis2010} and references therein.

\subsection{Modeling scenarios}
\label{sec:scenarios}
\begin{figure}[ht!]
\begin{center}
\includegraphics[angle=-90,width=0.47\textwidth]{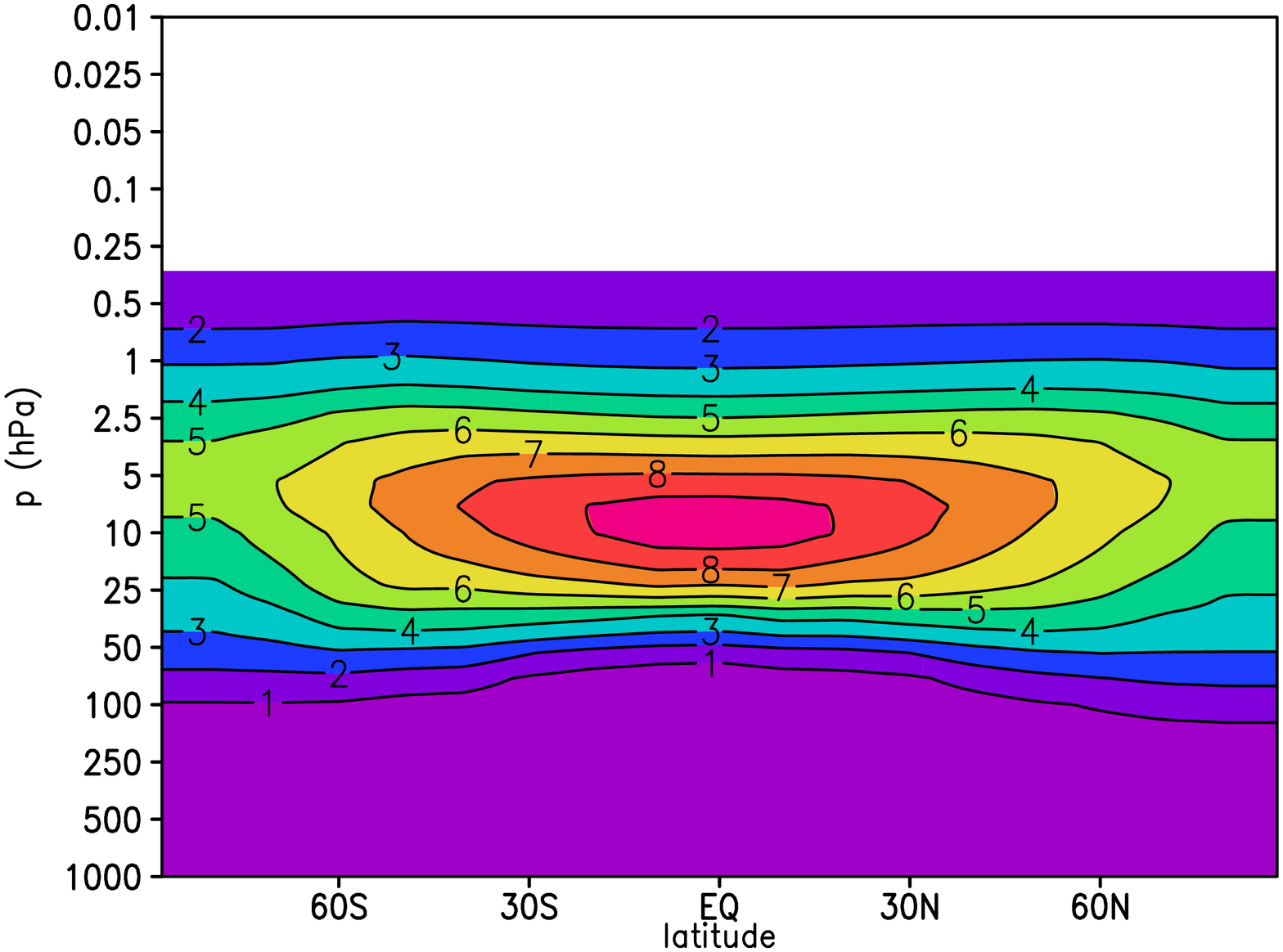}
\includegraphics[width=0.45\textwidth]{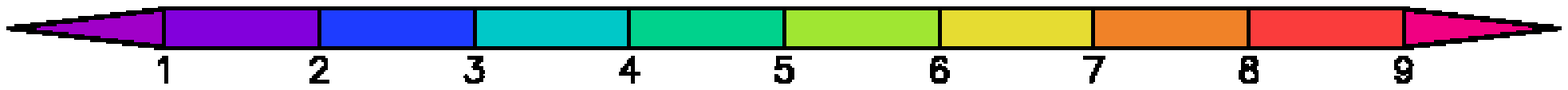}
\end{center}
\caption{Ozone concentrations (in ppmv) assumed in the 3D model calculations.}
\label{fig:ozone}
\end{figure}

\noindent
We investigate the influence of different main sequence stars upon the climate of Earth-like extrasolar planets. Planetary parameters are therefore chosen to resemble the present Earth, such as planetary radius, mass, hence gravity, obliquity, eccentricity, rotation rate, land-sea mask, and orography. Also an Earth-like atmospheric mass and chemical composition is assumed which is dominated by molecular nitrogen (N$_2$) and O$_2$ and includes spatially and temporally uniform trace gas amounts of CO$_2$ (\unit[355]{ppm}), CH$_4$ (\unit[1.64]{ppm}) and N$_2$O (\unit[308]{ppb}). For O$_3$ an annual mean of the zonal mean distribution given in Fortuin and Kelder (1998)\nocite{Fortuin1998} is taken, see Fig.~\ref{fig:ozone}.  We assumed present day O$_3$ concentrations, as O$_3$ has a large impact on the temperature structure and thereby on the dynamical processes in our present day atmosphere.  For the comparison 
of different relative humidities in section \ref{sec:1d3d} an atmosphere composed only of N$_2$, H$_2$O and CO$_2$ has been assumed.

In the 3D model the amount of water in the atmosphere, i.e.~water vapor, liquid water and water ice, is calculated considering various processes such as atmospheric transport, phase transitions within clouds and precipitation. In the 1D model the amount of water vapor is determined by the assumption of a relative humidity profile and the calculated temperature profile.\\  

\begin{figure}
 \centering
{\includegraphics[width=0.5\textwidth]{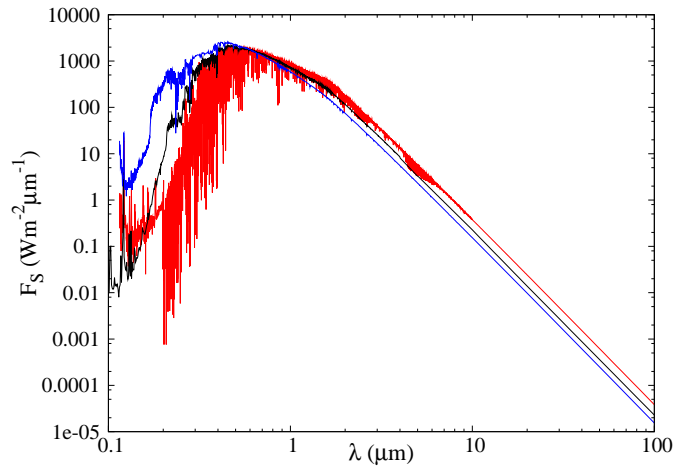}
 }
\caption{Stellar spectra (stellar energy flux $F_S$) used in this work.  Blue: F-type star, Black: Sun, Red: K-type star. The spectra all yield a total energy input of \unit[1366]{Wm$^{-2}$} at TOA, corresponding to the orbital distances given in table \ref{tab:scenario}.}
\label{fig:specs}
\end{figure}

In order to allow for ice free conditions in the 3D model calculations the background surface albedo of Antarctica and Greenland is set to 0.15 resembling the albedo of granite.  For the other glacier free surface area the albedo map from  \cite{Hagemann2002} is used.  The surface albedo considered by the radiative transfer is then calculated by the model taking the snow coverage of the land surfaces and canopy as well as the sea ice coverage and surface temperatures into account. However, no ice surfaces on land are calculated by the model, such as glaciers for instance. Therefore, the surface albedo for the reference scenario of the Earth around the Sun does not correspond to present Earth since snow on ice (e.g.~glaciers) has a larger albedo than snow on granite.  In the 1D model calculations the surface albedo is set to 0.2 (or 0.22 for an atmosphere composed of N$_2$, CO$_2$ and H$_2$O only, see sec.~\ref{sec:1d3d})
 which is the value needed to obtain the mean surface temperature of the Earth (\unit[288]{K}) for a solar spectrum with a Total Solar Irradiance (TSI) of \unit[1366]{Wm$^{-2}$} \citep{Gueymard2004}, to mimic the albedo effect of clouds.
Note that the assumed land-sea mask, the surface albedo, as well as the water reservoir have an impact on the model results. For planets with a small water reservoir, the humidity of the atmosphere would be reduced, and different land-sea masks would also change e.g.~the excitation of planetary waves, possibly leading to a different atmospheric circulation. The surface albedo is of great importance for the calculation of the radiative energy transport. In the present work, however we keep the number of parameters changed to a minimum as we are mainly interested in the impact of different stellar irradiations.
\\
For our central stars we use the Sun (G-type star) for the reference scenario, the F-type star $\sigma$ Bootis, and the K-type star $\epsilon$ Eridani (see table \ref{tab:scenario}). The stellar spectra are depicted in Fig.~\ref{fig:specs}. The orbital distances are chosen to yield a total energy input of \unit[1366]{Wm$^{-2}$} at the top of the atmosphere (TOA), which corresponds to the TSI of the present Sun. Since $\sigma$ Bootis is more luminous than the Sun this scaling of the total stellar energy flux corresponds to a larger orbital distance and thereby longer orbital period. For $\epsilon$ Eridani, which is less luminous, the opposite is the case.  The orbital periods have been determined  using Kepler's 3rd law and the stellar masses given in table \ref{tab:scenario}. For these types of central stars and the relatively large orbital distances, one may reasonably assume that the planets are not forced into synchronous rotation by tidal forces \citep{Kasting1993hz}, therefore a rotation rate of the present Earth may be assumed. Table \ref{tab:scenario} summarizes the stellar properties as well as the orbital distances and periods of the planets. Details of the stellar spectra, which are a composite of satellite measurements in the UV and synthetic model spectra at larger wavelengths can be found in \cite{Kitzmann2010}. The solar spectrum is taken from \cite{Gueymard2004}. \\
The mixed layer ocean model which is coupled to the 3D  atmospheric model requires a flux correction, the so-called q-flux, to account for oceanic heat transport in the calculation of the sea surface temperatures. For Earth climate calculations this heat flux is usually calculated from a reference scenario with prescribed sea surface temperatures (SSTs) and then applied to modeling scenarios with a small disturbance of the reference state. It therefore depends on the model and model setup such as horizontal resolution. For large deviations from the reference state, a full atmo\-sphere-ocean GCM is usually applied, e.g. for long-term Earth climate predictions. The utilization of a simple mixed-layer ocean model was preferred to a coupled at\-mo\-sphere-ocean circulation model due to computational cost and the desire to limit the number of unknown boundary conditions. We therefore assume that the difference in the stellar spectral flux distribution to be a small  disturbance of the reference state. The q-flux is calculated from net surface heat fluxes ($F_{heat}$), i.e.~net radiative, sensible and latent heat flux, which have been calculated for a reference scenario of the present Earth using prescribed climatological monthly mean sea surface temperatures ($T_{SST}$) from  AMIP II (Atmospheric Model Intercomparison Project II, Taylor et al., 2000\nocite{AMIPII}) via:\\

\begin{equation}
q=F_{heat}-C_m\frac{\partial T_{SST}}{\partial t} \text{,}
\end{equation}

with $C_m$ the heat capacity of the ocean.
We make use of different q-flux corrections: 
\begin{itemize}
\item q1: the q-flux varies with every time step ($\partial t$). It uses monthly mean (mm) values of $F_{heat}$, which are calculated from a reference scenario, and $\frac{\partial T_{SST}}{\partial t}$, which varies with every time step and uses prescribed SSTs . This is the q-flux which would be applied for Earth climate calculations, as with this parametrization the prescribed SSTs of the reference scenario can be reproduced.
\end{itemize}
For the calculation of planetary climates at different orbital periods the above is however not useful, since it assumes a seasonality of a 365 days orbit. Therefore, different q-fluxes have been applied and their influence is discussed in section \ref{sec:sensitivity}.
\begin{itemize}
 \item q2: the q-flux also varies with every time step but it uses the annual mean (am) of $F_{heat}$, and  $\frac{\partial T_{SST}}{\partial t}$ varies with every time step,  
 \item q3: the q-flux is the monthly mean (mm) of q1, 
 \item q4: the q-flux is the annual mean (am) of q1, and 
 \item q5: the q-flux is equal to zero.
\end{itemize}

The q-fluxes q2, q3, and q4 represent only a small deviation from the usually applied q1, mainly affecting the temporal consideration of the oceanic heat flux. Assuming the q5 q-flux completely ignores any heat transport. The q-fluxes q2-q4 test wh\-ther a small change in the oceanic heat flux can lead to large deviations of the climate stages, as expected for Earth climate simulations. The q-flux q5 tests the robustness of our results.

Note that for the results discussed in sections \ref{sec:temperatures}--\ref{sec:surf} q2 is used for the planet around the K-type star and q4 is used for the planet around the F-type star. The difference in the annual global mean surface temperature owing to the different q-fluxes is however small as shown in section \ref{sec:sensitivity}.\\

The seasons have different lengths depending on the orbital period. The length of the seasons, as e.g.~northern hemispheric winter (NHW, used in sec.~\ref{sec:temperatures} and \ref{sec:hydrocycle}), has been determined by the distribution of stellar insolation over latitudes and time. Hence, for each scenario the atmospheric temperature has been averaged over a time period for which the stellar insolation over latitudes corresponds to e.g.~December, January and February for the Earth around the Sun. For the planet around the K-type star a NHW lasts \unit[44]{days} and to \unit[212]{days} for the planet around the F-type star. Furthermore, the results have been averaged over several orbits for which the atmosphere has reached a quasi-equilibrium state (after spin-up, when the influence of the initial state (e.g.~stellar insolation) has ceased and surface temperature variations are caused by the seasonal cycle only): 6 orbits for the planet around the F-type star, 26 orbits for the planet around the Sun, and 18 orbits for the planet 
around the K-Star. 
For the planet around the F-type star this corresponds to more than 5000 days due to the longer orbital period (and only to about 3000 days for the planet around the K-type star). We limited the number of orbits included in the long term mean to save computing time.\\ 
In section \ref{sec:sensitivity} the influence of different orbital periods, and hence different durations of seasons is discussed briefly. The orbital period has been varied because in principle different lengths of seasons may lead e.g.~to different sea surface temperatures, because the ocean has a large thermal inertia. In addition, stellar parameters such as e.g.~the stellar mass, which is used to calculate the orbital periods, also has uncertainties. This partly motivates the assumption of different various orbital periods.\\

\begin{table}
\begin{center}
\begin{threeparttable}[ht]
\centering
\caption{Stellar and orbital parameters.}
\label{tab:scenario}

\centering
\begin{tabular}{lllll}
Star&stellar type&%T$_{eff}$ (K)& 
M/M$_{Sun}$&  $a$/au &$P$/days\\
\hline\hline
$\sigma$ Bootis & F2V& %6722\tnote{a}&
1.194\tnote{a}&1.89& 868.64\\
Sun & G2V& %5777\tnote{c}&
1&1& 365.25\\
$\epsilon$ Eridani&K2V&%5073\tnote{d}&
0.82\tnote{b}&0.6& 184.00\\
\hline
\end{tabular}
%\end{center}
\begin{tablenotes}
%\item[a] \cite{Cenarro2007} 
\item[a]\cite{Boyajian2012} 
%\item[c] \cite{Cox2000} \item[d] \cite{Santos2004} 
\item[b] \cite{Butler2006}
\end{tablenotes}
\end{threeparttable}
\end{center}
\end{table}

\section{Results and discussion}
\label{sec:results}
\noindent
To investigate the consequence of the different input spectra upon the climate we first discuss the results of the 3D model calculations for the Earth-like extrasolar planets orbiting different types of main-sequence stars in terms of surface temperatures, temperature structures, hydrological cycle and surface properties, and then compare these results to those of the 1D model. 

\subsection{Temperatures}
\label{sec:temperatures}
\begin{figure}[ht!]
 \centering
{\includegraphics[angle=-90,width=0.4\textwidth]{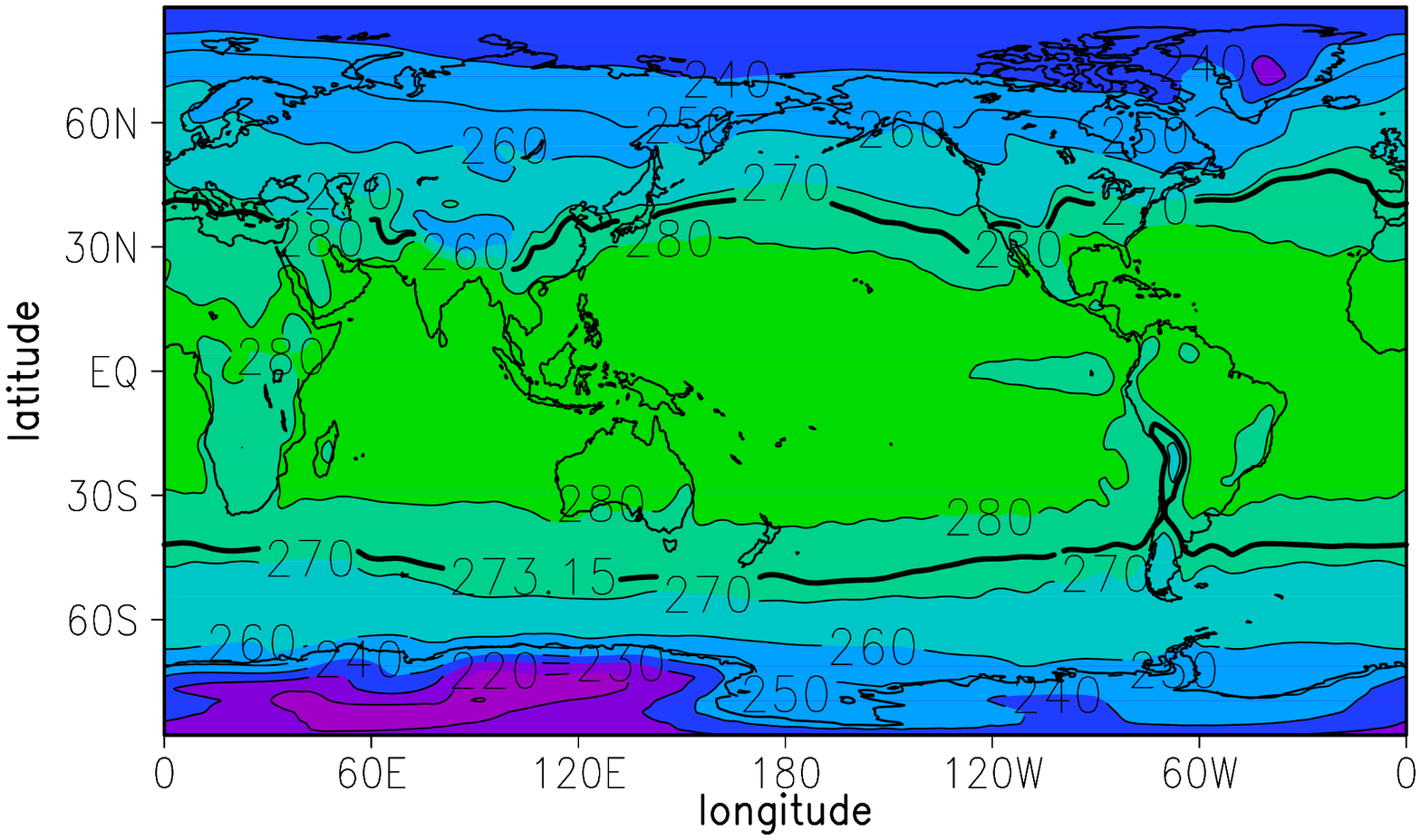} 
\includegraphics[angle=-90,width=0.4\textwidth]{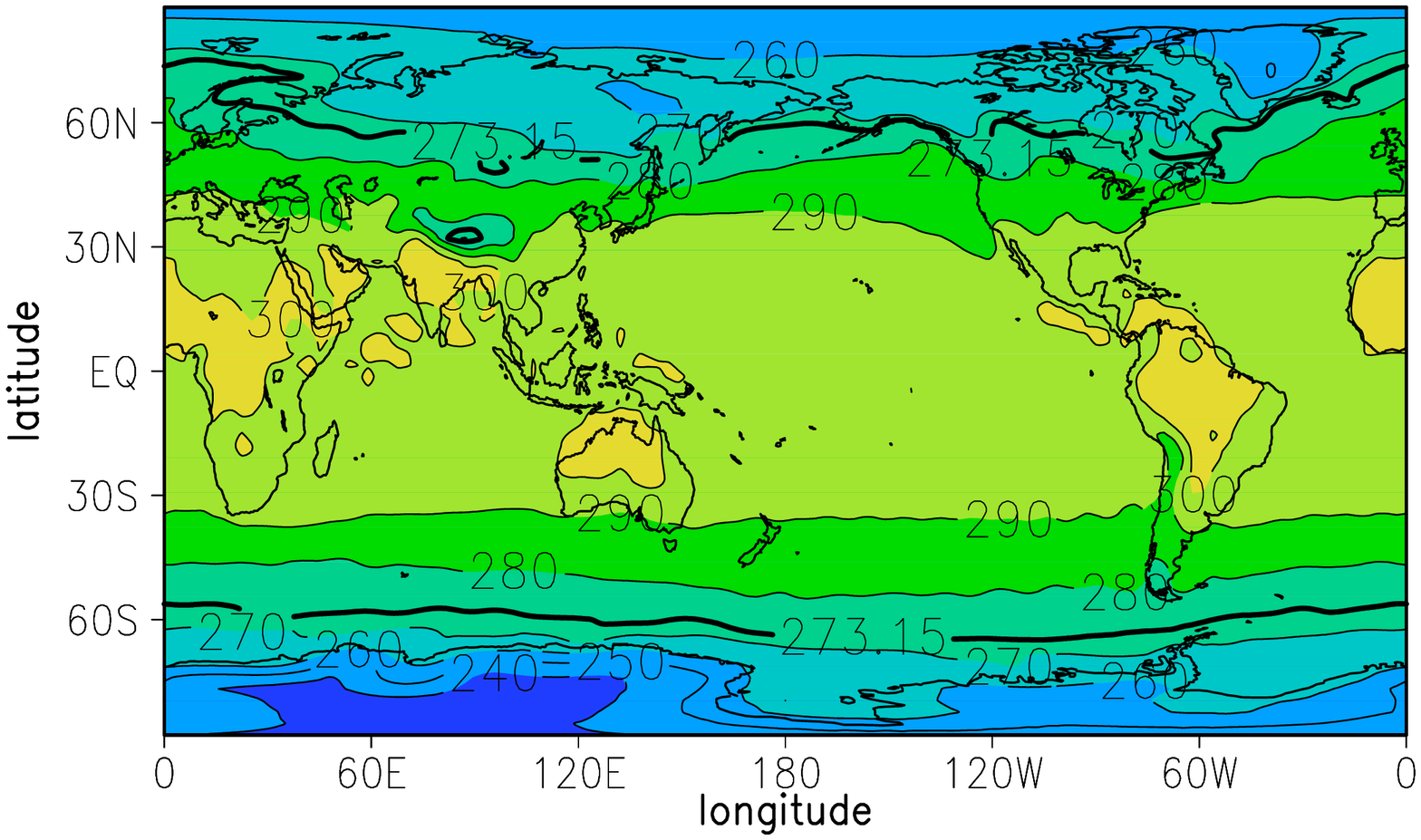}
\includegraphics[angle=-90,width=0.4\textwidth]{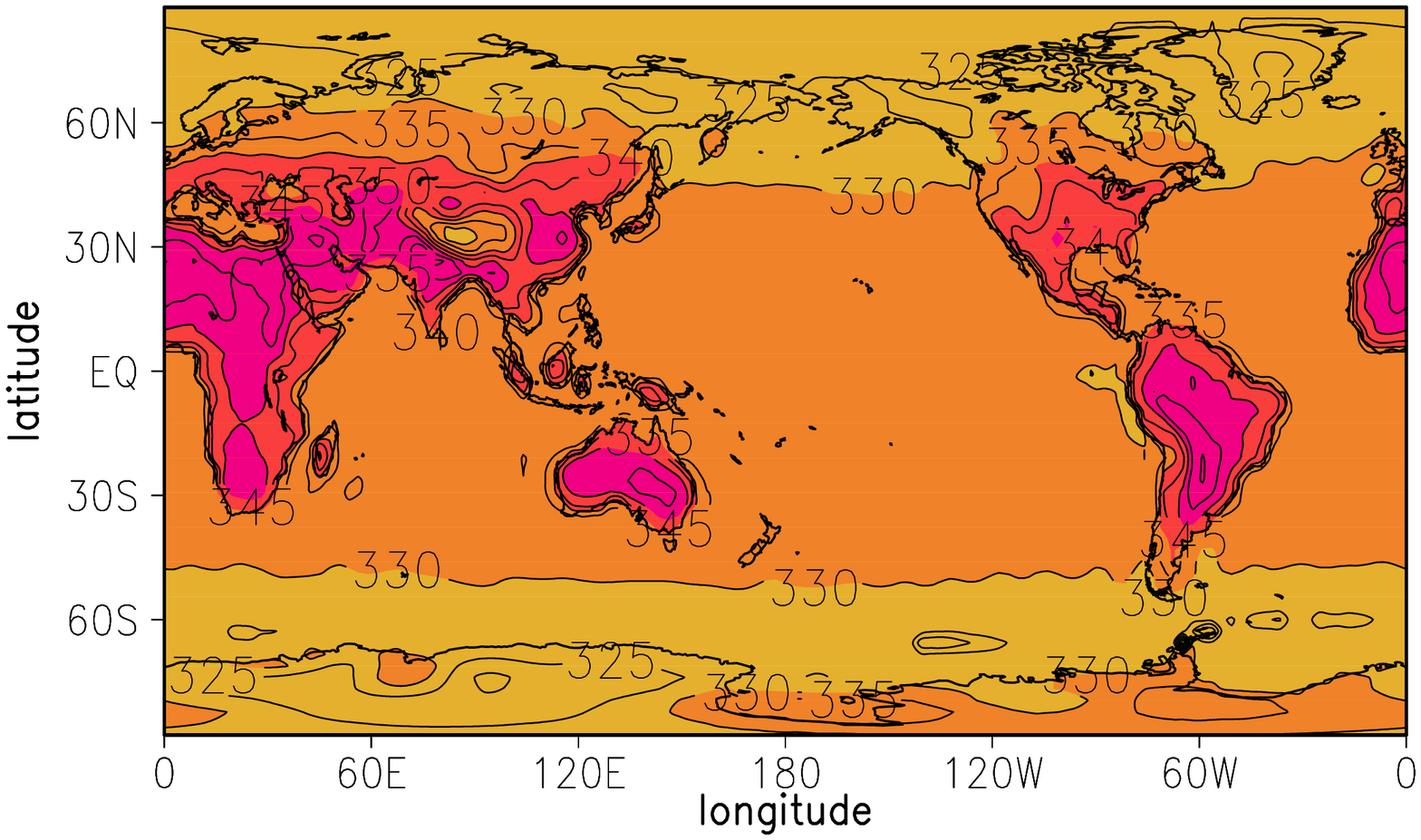}
\includegraphics[angle=-90,width=0.5\textwidth]{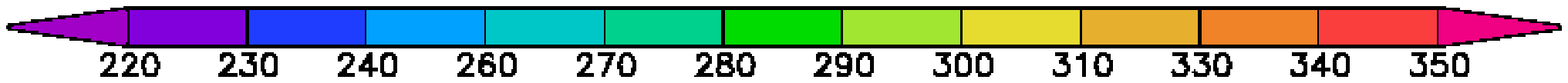}
}
\caption{Orbital mean near surface temperatures (K) for Earth-like planets around the F-type star (upper panel), the Sun (middle panel) and the K-type star (lower panel). The thick black line corresponds to the melting temperature of water (\unit[273.15]{K})}
\label{fig:2mT}
\end{figure}

\noindent
For the question of a planet's habitability the temperatures at the surface are relevant. In the following first the response of the 2-me\-ter-tem\-pe\-ra\-ture (in the forthcoming text termed 'near-surface temperature'), i.e.~the atmospheric temperature in the lowermost atmospheric layer, hence closest to the surface is discussed.  
Figure \ref{fig:2mT} depicts the orbital mean of the near-surface temperature for the Earth-like planets around different types of stars over latitudes and longitudes.
Despite the fact that all three planetary scenarios receive the same amount of stellar energy, the near-surface temperatures are quite different. 
Temperatures are lower for the planet around the F-type star and higher for the planet around the K-type star in comparison to the Earth-like planet around the Sun. \\

For the planet around the F-type star large regions of the planetary surface are ice free and habitable, with temperatures well above the freezing point of water (\unit[273.15]{K}) despite a global mean temperature of \unit[273.6]{K}, see table \ref{tab:gm_orb}.
Temperatures below the freezing point of water are found polewards of about 40$^{\circ}$ latitude. For Early Earth, e.g.~\cite{Kunze2014}, \cite{Charnay2013}, Wolf and Toon (2013)\nocite{Wolf2013} have also shown that global mean temperatures below the freezing point of water do not necessarily lead to freezing of the entire surface reservoir of liquid water, to so-called snowball states. Instead, regions of open water in the equatorial region may be found even for global mean temperatures as low as \unit[250]{K}. This should be kept in mind when 1D model results are used to evaluate the habitability of an extrasolar planet, Early Earth or Early Mars.\\

For the planet around the K-type star the near surface temperature is everywhere higher than the maximum temperature obtained for the planet around the Sun,  i.e.~temperatures at polar latitudes for the planet around the K-type star are larger than equatorial temperatures for the planet around the Sun. 
The global orbital mean near surface temperatures (T$_{2m}$) for the scenarios are summarized in table \ref{tab:gm_orb}.\\  

\begin{table*}
\caption{Global mean atmospheric and surface response for planets around different types of central stars.}
\begin{center}
\begin{tabular}{lccc}

\hline
 & F3D & Sun3D & K3D\\\hline
Star (Stellar type) & $\sigma$ Boo (F2V) & Sun (G2V)& $\epsilon$ Eri (K2V)\\
T$_{2m}$ (K) & 273.6 & 288.6 & 334.9\\
GHE (K) & 20.8 & 39.2 & 73.0\\
Planetary albedo & 0.38 & 0.36 & 0.23\\
Water vapor column (\unit[]{$\tfrac{kg}{m^2}$}) & 10.2 &  30.1 & 482.6 \\
Cloud cover (\%) & 0.75& 0.70&0.67\\
Cloud water column (\unit[]{$\tfrac{kg}{m^2}$}) & 0.103 & 0.106 & 0.203\\
Cloud ice column (\unit[]{$\tfrac{kg}{m^2}$}) &0.037& 0.029 & 0.028  \\
Cloud GHE (K) & 9.0 & 8.9& 11.0\\
Surface albedo & 0.21& 0.15 & 0.10 \\
Sea ice fraction relative to ocean (\%)  &11.4 & 3.9& 0.0\\
Snow depth (m) & 0.0109 &0.0039 & 0.0 \\

\end{tabular}
\label{tab:gm_orb}
\end{center}
\end{table*}

These temperatures are the result of the interaction of various processes, such as absorption and scattering of stellar radiation by the atmosphere and surface, the greenhouse effect, and energy transport by convection and dynamics. Our results confirm earlier studies (e.g.~Kasting et al., 1993, Segura et al., 2003, Kitzmann et al., 2010, Shields et al., 2013\nocite{Kasting1993hz, Segura2003, Kitzmann2010, Shields2013}) which already discussed, that the spectral distribution of the stellar energy leads to different temperature responses of the atmosphere. The F-type star has its radiation maximum at shorter wavelengths and the K-type star at longer wavelengths than the Sun (see Fig.~\ref{fig:specs}). Therefore, radiation from the F-type star is more effectively scattered back to space via Rayleigh scattering, while radiation by the K-type star is less effectively scattered and more effectively absorbed by water vapor, and carbon dioxide in the NIR. This leads to a higher planetary albedo for the planet around the F-type star and a lower planetary albedo for the planet around the K-type star (see table \ref{tab:gm_orb}). The planetary albedo has been determined from the outgoing longwave radiation assuming energy balance at the top of the atmosphere. 
Hence, the net incoming radiation (incident radiation -- backscattered radiation) is thereby smaller for the planet around the F-type star and larger for the planet around the K-type star compared to a planet around the Sun, although the total top-of-atmosphere (TOA)  incident radiation flux is the same for all three cases. Furthermore more stellar radiation is absorbed by H$_2$O and CO$_2$  for the planet around the K-type star, and less by the planet around the F-type star, respectively. This results in lower temperatures for the planet around the F-type star and in higher temperatures for the planet around the K-type star, as shown in Fig.~\ref{fig:2mT}. 

This temperature response is intensified by positive climate feedbacks, such as the water vapor feedback cycle or the ice-albedo feedback which strongly depend on local circumstances, such as temperatures, but have an impact on the global climate state. 

\begin{figure}[ht!]
 \centering
{
\includegraphics[angle=-90,width=0.4\textwidth]{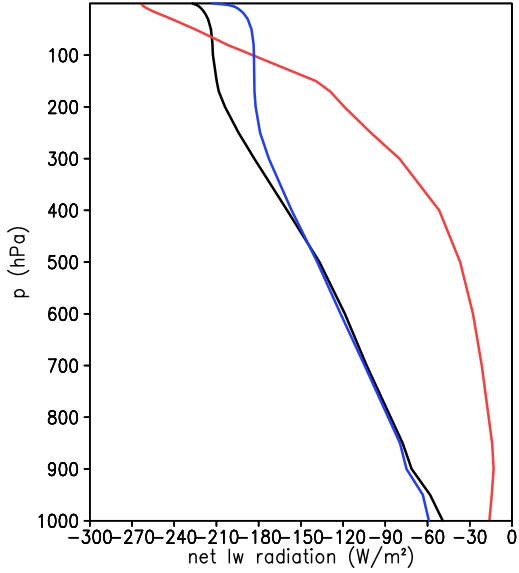}
}
\caption{Global orbital mean net longwave radiation, $F_{lw,net}=F_{lw,down}-F_{lw,up}$, for the planet around the F-type star (blue), planet around the Sun (black), planet around the K-type star (red). The negative values indicate the upward direction.}
\label{fig:Ftherm}
\end{figure}

For the planet around the F-type star the lower surface temperatures yield lower water vapor concentrations, which is discussed in section \ref{sec:hydrocycle}, see Fig.~\ref{fig:h2o}. The water vapor column is given in table \ref{tab:gm_orb}. The lower water vapor concentrations lead to a decrease in the greenhouse effect. For the planet around the K-type star the opposite is the case, i.e.~higher water vapor concentrations and a larger greenhouse effect. The greenhouse effect  (GHE) in terms of temperature (given in table \ref{tab:gm_orb}) has been inferred from the difference of the temperature corresponding to the upwelling longwave radiation at the surface and the  temperature corresponding to upwelling longwave radiation at TOA via the Stefan-Boltzmann law. The GHE for the planet around the Sun is a little higher than that of the Earth (about \unit[33]{K}), which is probably caused by the assumption of the low surface albedo map leading to little higher temperatures and hence also higher water vapour concentrations. The strength of the greenhouse effect is illustrated in terms of the net  longwave radiation ($F_{lw,net}$) close to the surface  in Fig.~\ref{fig:Ftherm}. $F_{lw,net}$ is the difference of the upward and downward directed longwave radiation ($F_{lw,net}=F_{lw,down}-F_{lw,up}$), hence negative values correspond to upward directed, positive values to downward directed radiation, respectively. Values close to zero, as for the planet around the K-type star, correspond to a large greenhouse effect, because a large amount of the upward directed radiation is absorbed and re-emitted by the atmosphere leading to less negative values. More negative numbers close to the surface, as for the planet around the F-type star, correspond to a weaker greenhouse effect because the upward directed radiation is less effectively absorbed, and thereby the net longwave radiation is more negative. 
At larger heights the planet around the K-type star shows a more negative $F_{lw,net}$, due to the higher atmospheric temperatures, and the planet around the F-type star less negative longwave radiation. \\

\begin{figure}[ht!]
 \centering
{

\includegraphics[angle=0,width=0.23\textwidth]{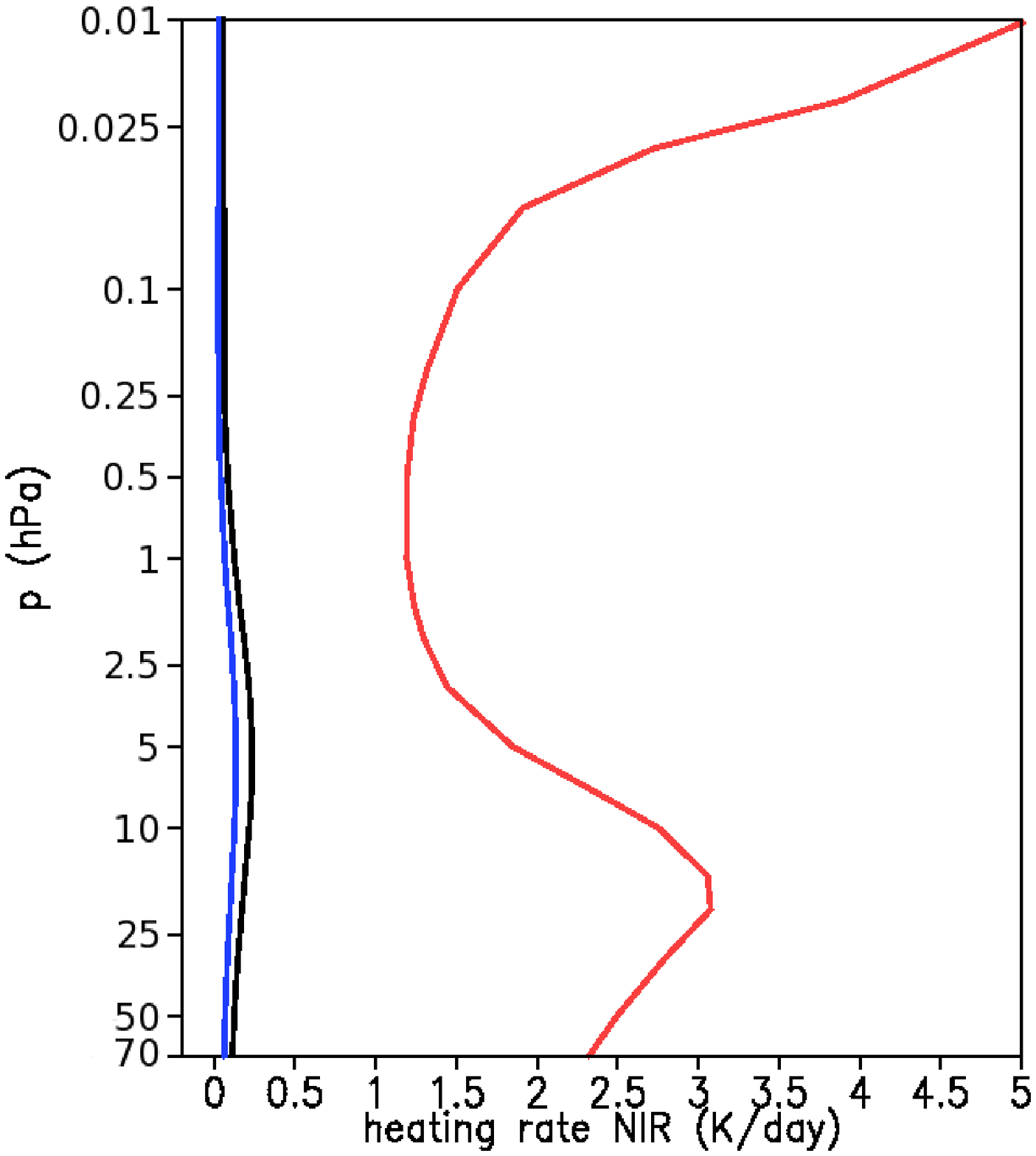}
\includegraphics[angle=0,width=0.23\textwidth]{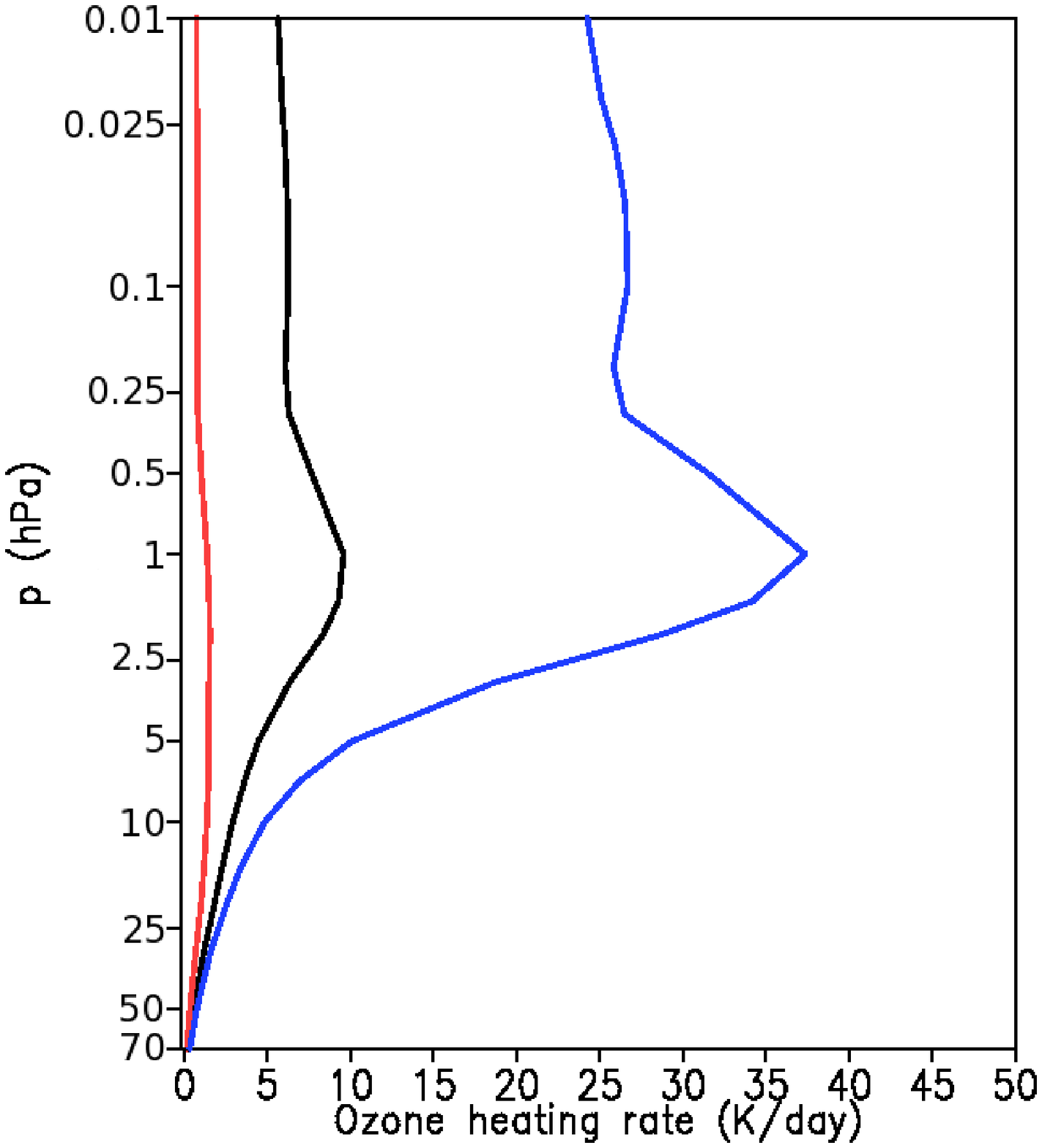}
}
\caption{Global mean shortwave heating rates (for p$\le$\unit[70]{hPa}) in the NIR (left) and due to O$_3$ (right) for the planet around the F-type star (blue), planet around the Sun (black), planet around the K-type star (red).}
\label{fig:heating}
\end{figure}

The increase in temperature in the lower atmosphere for the planet around the K-type star is amplified by a coupling of the stellar radiation flux to the water vapor feedback cycle. Due to the SED of the K-type star stronger radiative heating occurs in the lower atmosphere caused by the absorption of NIR radiation by water vapor, carbon dioxide and clouds, as can be seen in the right panel of Fig.~\ref{fig:heating}. This results in higher tropospheric and surface temperatures, leading to a higher amount of water vapor in the atmosphere by evaporation (see Fig.~\ref{fig:h2o}). This increase in water vapor enhances the greenhouse effect (see also Fig.~\ref{fig:Ftherm}) and the absorption of stellar radiation in the NIR, consequently yielding higher surface temperatures. This eventually leads to melting of all the ice and snow cover lowering the surface albedo (as discussed in section \ref{sec:surf}, shown in Fig.~\ref{fig:surf}), which 
results in less reflection of stellar light at the planetary surface, hence more absorption, causing even higher surface and tropospheric temperatures.\\

\begin{figure}[ht!]
 \centering
{
\includegraphics[angle=-90,width=0.4\textwidth]{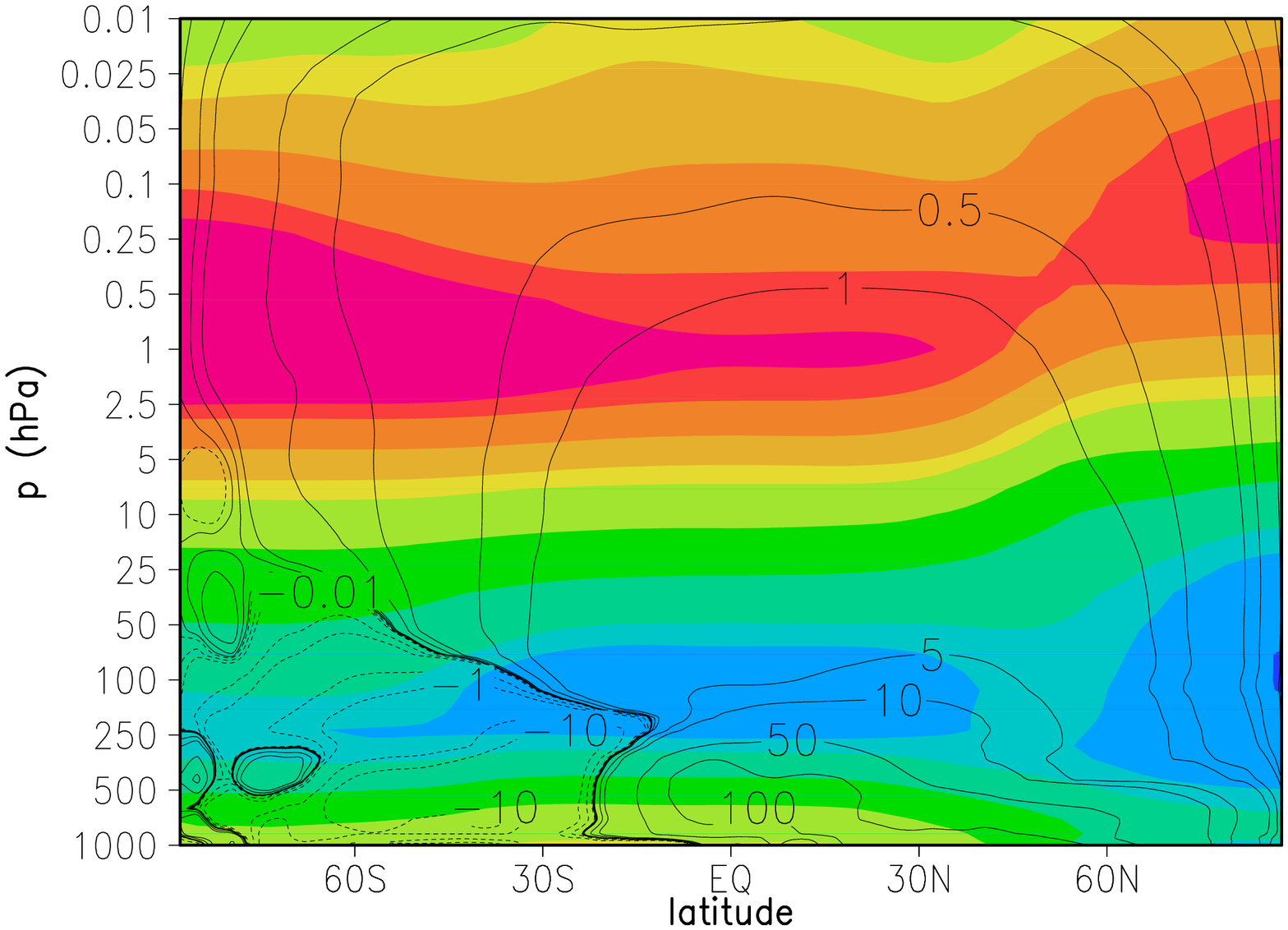}
\includegraphics[angle=-90,width=0.4\textwidth]{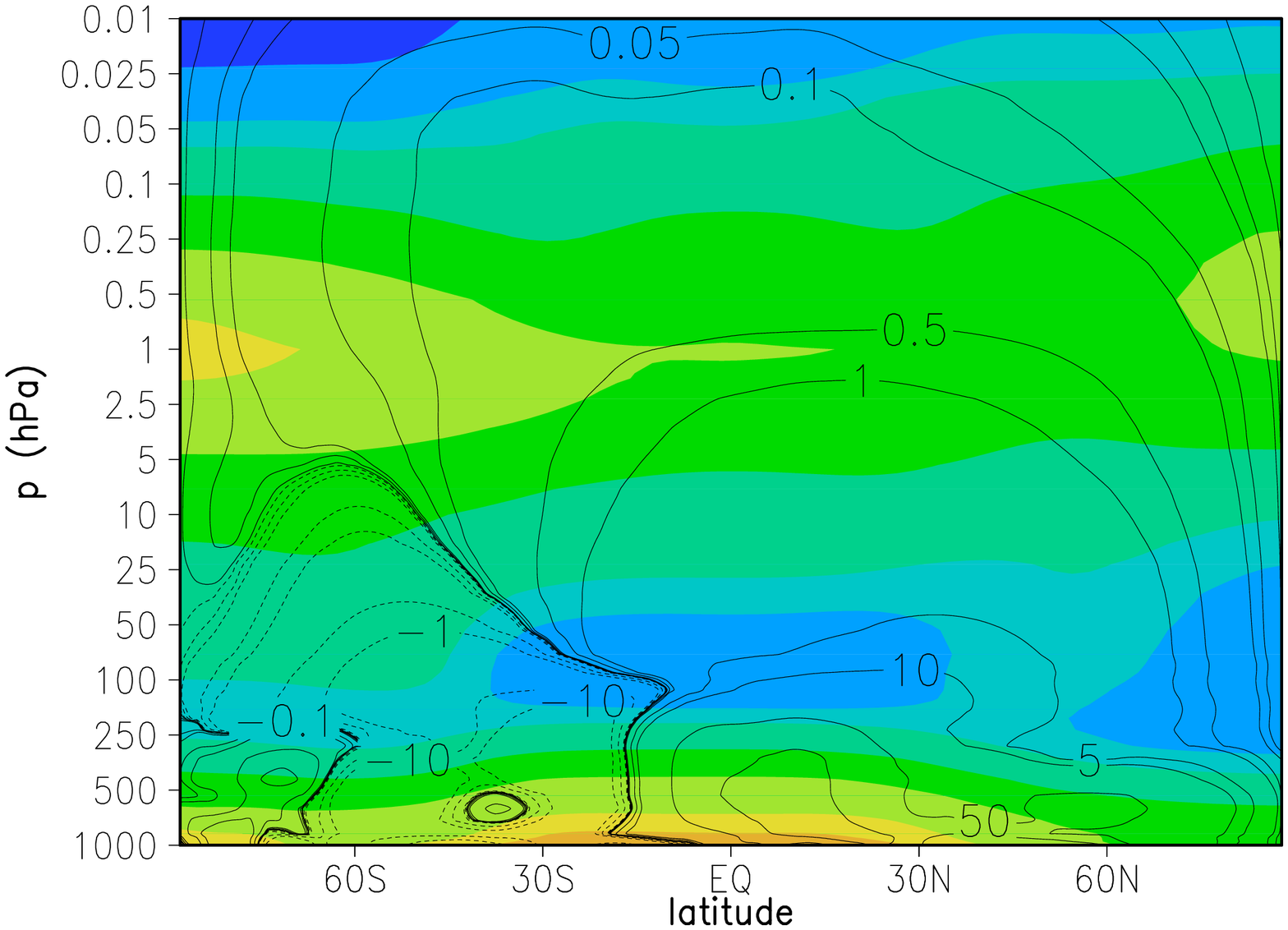}
\includegraphics[angle=-90,width=0.4\textwidth]{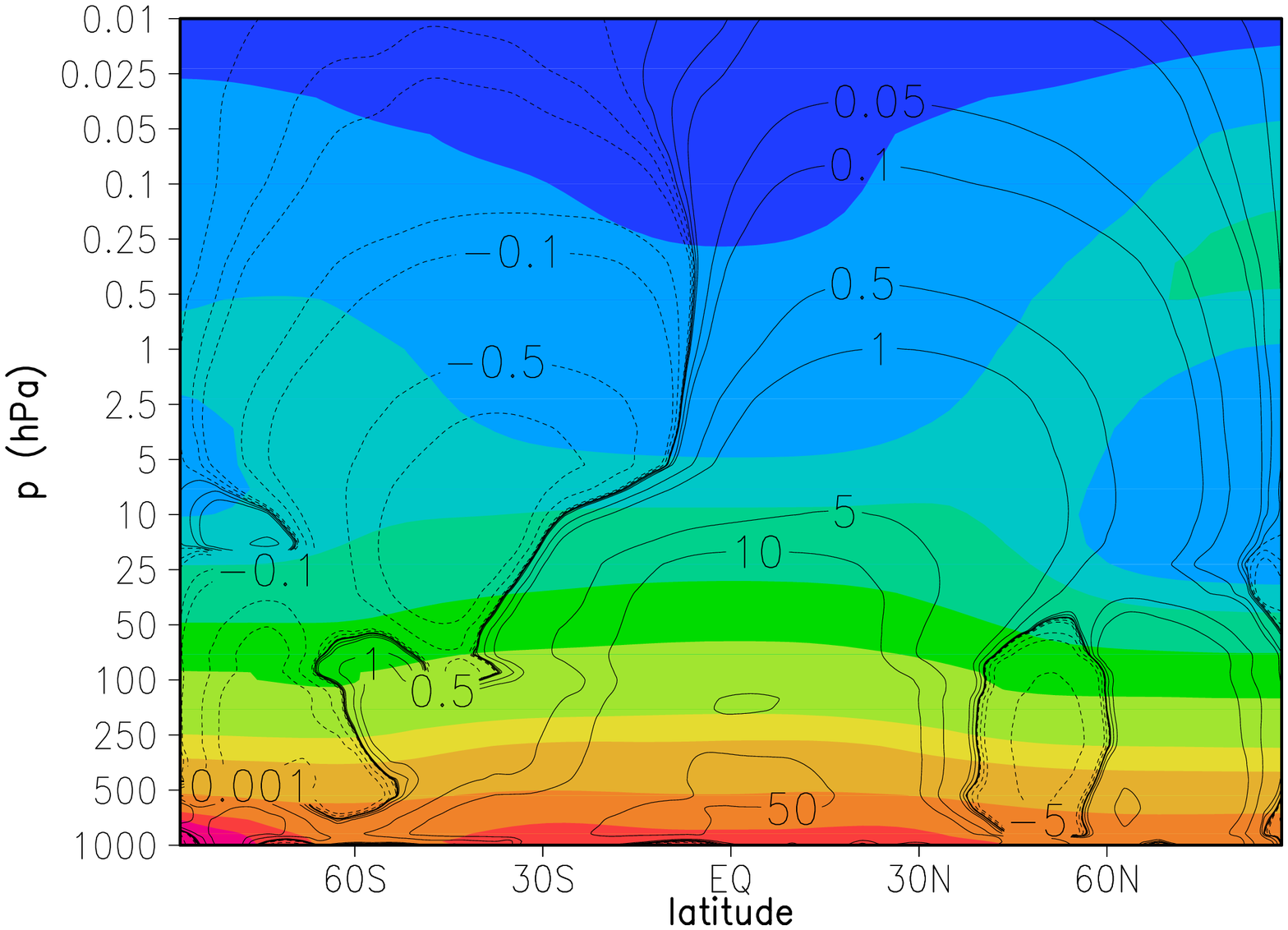}
\includegraphics[angle=-90,width=0.5\textwidth]{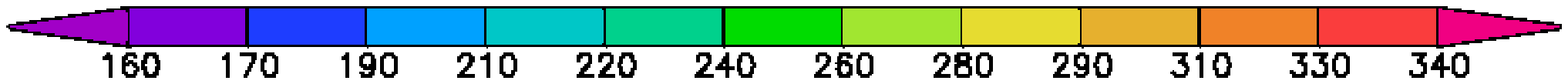}
}
\caption{Zonal mean atmospheric temperature (K) structure for northern hemispheric winter (NHW) for planets around different stars. Upper panel: planet around the F-type star, middle panel: planet around the Sun, lower panel: planet around the K-type star. The black contour lines show the meridional residual mass stream function in \unit[10$^9$]{kgm$^{-2}$}, where solid lines represent clockwise, dotted lines anti-clockwise circulation.}
\label{fig:Tstruc}
\end{figure}

The different SEDs, however, not only result in a change of the surface conditions, but also affect the vertical temperature structure of the atmosphere. Figure \ref{fig:Tstruc} shows the zonal mean temperature structure for northern hemispheric winter (NHW, which corresponds to a December-January-February mean for the planet around the Sun) for all three scenarios from the surface up to a top pressure of \unit[0.01]{hPa}.
The zonal mean temperature structure for the southern hemispheric winter is not shown since the temperature differences between the hemispheres are small compared to the response to the different stellar spectral distributions.\\

For the reference scenario of the planet around the Sun (middle panel of Fig.~\ref{fig:Tstruc}) the typical vertical temperature structure of the Earth is visible with highest temperatures at the surface. Temperatures decline throughout the troposphere, reaching a minimum at the tropopause. Within the stratosphere temperatures increase reaching a maximum in the stratopause and decrease again in the mesosphere. The zonal temperature structure shows highest tropospheric temperatures in the equatorial region, and larger temperatures in the summer than in the winter hemisphere. Note that the resulting tropospheric temperatures at the summer south pole are higher than for the Earth, which is a result of the low background albedo map, see sec.~\ref{sec:scenarios}. The tropopause is coldest in the equatorial region and during polar night.
At the stratopause highest temperatures occur during polar day due to the absorption of stellar radiation by O$_3$. Similar temperatures at the stratopause are also found for polar night. These are the result of the meridional circulation, also called Brewer-Dobson-Circulation, which is indicated by the black contour lines in Fig.~\ref{fig:Tstruc} (solid lines show clockwise, dashed lines anti-clockwise circulation, respectively). Air is lifted from the equatorial tropopause through the summer stratosphere and mesosphere, and then transported to the winter hemisphere where it sinks. During ascent the air cools adiabatically, causing a cold summer mesosphere, and during descent the air heats adiabatically, causing the temperature maximum at the winter stratopause.\\

For the planet around the F-type star (upper panel of Fig.~\ref{fig:Tstruc}), the overall temperature structure is similar to that of the Earth. However, tropospheric temperatures are lower than for the solar spectrum, whereas stratospheric temperatures are much higher, reaching maximum values of more than \unit[350]{K}, i.e.~larger than the surface temperature. The high stratospheric temperatures in the summer hemisphere are caused by the high amount of stellar radiation at wavelengths where ozone absorbs, which is clearly visible in the ozone heating rates shown in the right panel of Fig.~\ref{fig:heating}. The stratopause temperatures in the winter hemisphere are similar to those of the summer hemisphere. They result from the stratospheric meridional circulation, which is illustrated by the black contour lines in Fig.~\ref{fig:Tstruc}. 

For the planet around the K-type star (lower panel of Fig.~\ref{fig:Tstruc}) the overall temperature structure differs considerably from that of the reference scenario around the Sun. There is no pronounced temperature increase within the stratosphere unlike the other two scenarios, because the stellar spectrum features much reduced radiation in the wavelength region where absorption by ozone occurs (at about \unit[200-800]{nm}) compared to the Earth around the Sun, which results in less ozone heating (right panel of Fig.~\ref{fig:heating}). Furthermore, the pronounced cold tropopause in the equatorial region is missing. At lower pressures, however, atmospheric temperatures become very low (below \unit[190]{K}) consistent with adiabatic cooling of air during ascent. 

Due to the high surface temperatures the troposphere expands and shows nearly constant temperatures over all latitudes up to a pressure of \unit[50]{hPa}. At pressures of about \unit[5]{hPa} features similar to the cold lower stratosphere of the Earth appear in both polar regions. Despite the fact that no pronounced stratospheric temperature maximum exists in the summer hemisphere, a temperature maximum in the winter stratosphere is evident. This results from meridional circulation in the upper atmosphere from the equator to the poles (indicated by the black contour lines), which leads to a warming of the upper winter hemisphere due to adiabatic warming of descending air parcels. Note, that the meridional circulation patterns in the stratosphere are different for the planet around the K-type star compared to the planet around the Sun and the F-type star. The air rises from the equatorial troposphere through the equatorial stratosphere and mesosphere and sinks in both polar regions. This is in contrast to the Brewer-Dobson-Circulation where the air rises through the summer stratosphere and mesosphere and sinks in the winter hemisphere. 

\begin{figure}[ht!]
 \centering
{
\includegraphics[angle=-90,width=0.43\textwidth]{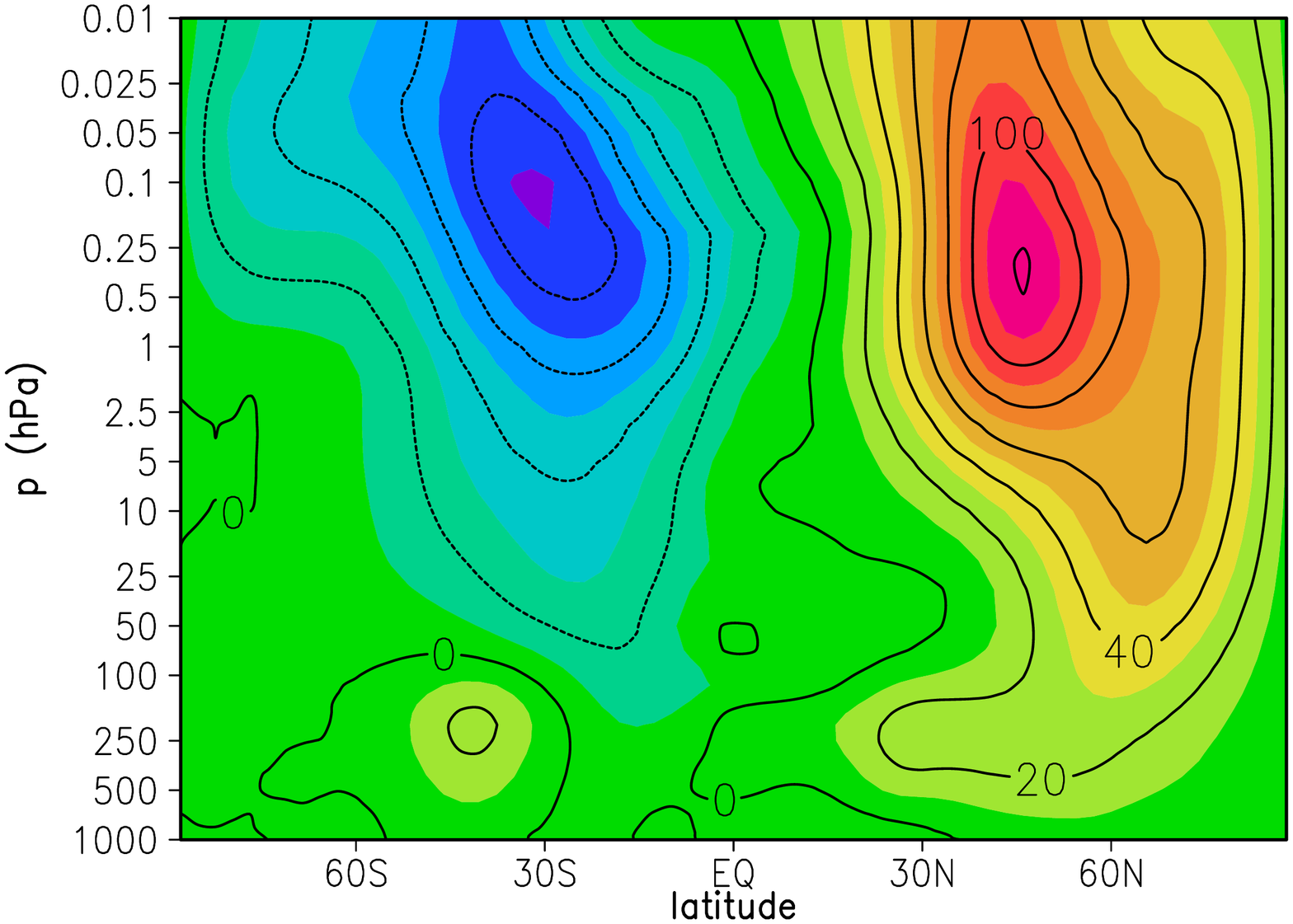}
\includegraphics[angle=-90,width=0.43\textwidth]{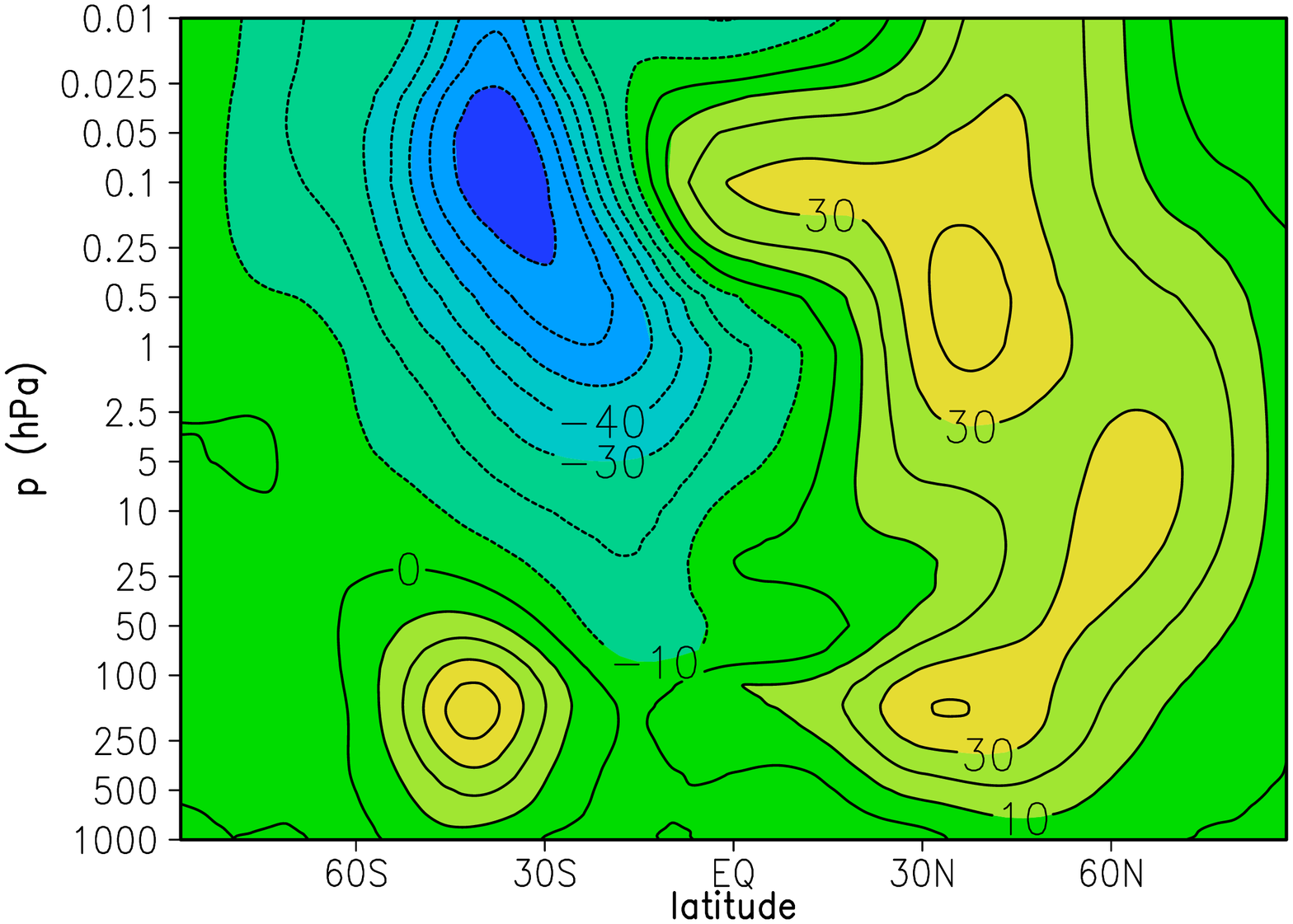}
\includegraphics[angle=-90,width=0.43\textwidth]{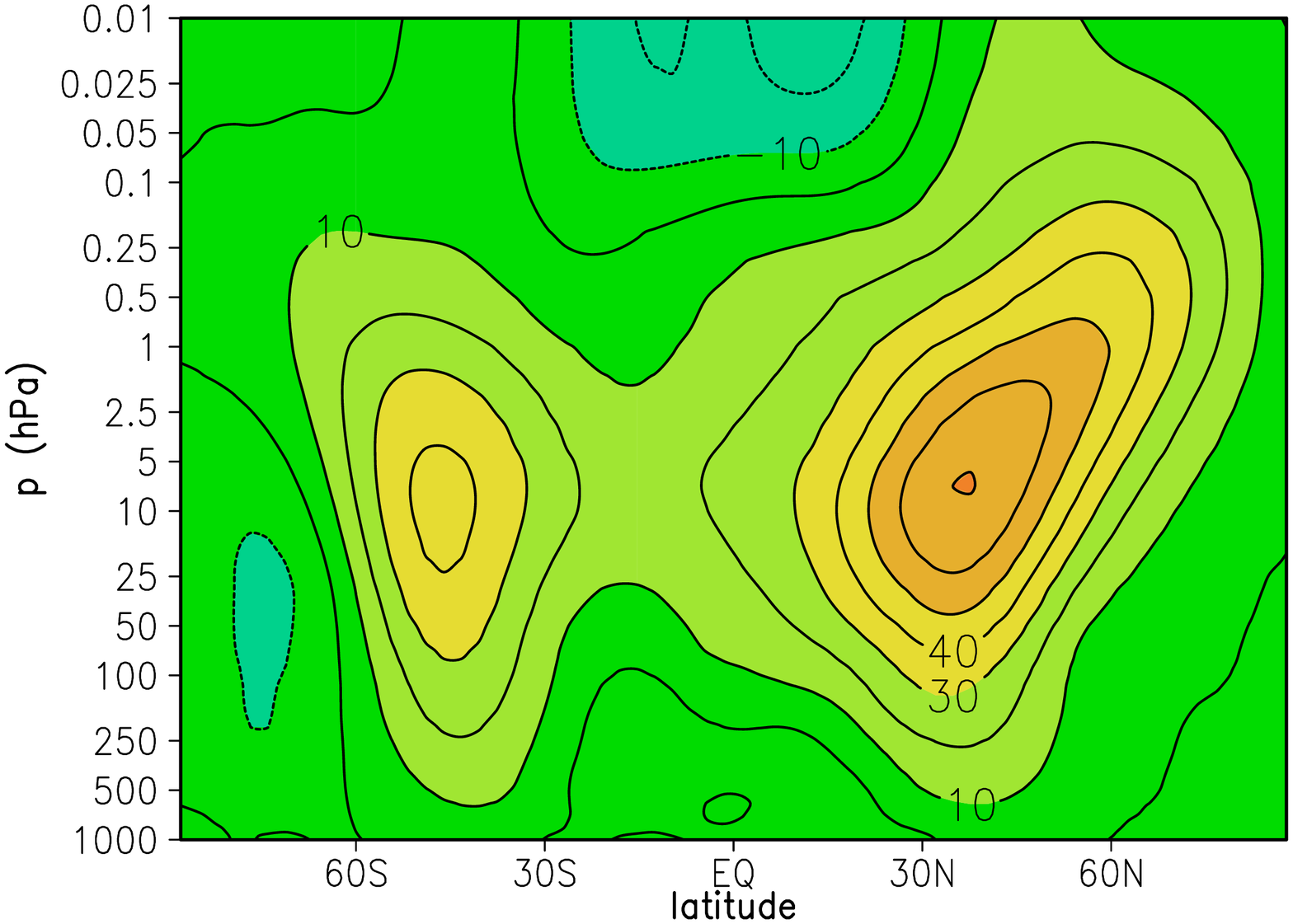}
\includegraphics[angle=0,width=0.5\textwidth]{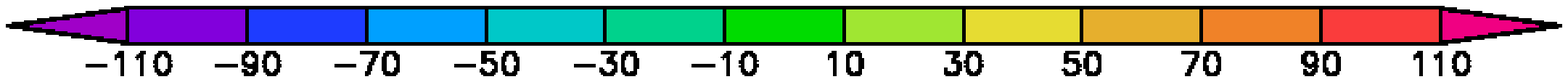}
}
\caption{Zonal mean wind (m/s) structure for northern hemispheric winter (NHW) for planets around different stars. Upper panel: planet around the F-type star, middle panel: planet around the Sun, lower panel: planet around the K-type star. Positive values indicate westerlies, while negative values show easterlies.}
\label{fig:Ustruc}
\end{figure}

This change in the meridional circulation is a result of the different temperature structures which leads to a strong change in the zonal wind structure as shown in Fig.~\ref{fig:Ustruc}. 
For all scenarios we find westerly winds in the troposphere. These are caused by the temperature gradient from the equator to the pole. The warm air flows poleward and is deflected to the east due to the Coriolis force. For the planet around the F-type star the westerly winds are weaker than for the planet around the Sun due to the smaller tropospheric temperature gradient. 
The stratospheres for the planets around the F-type star and the Sun show a westerly wind in the winter hemisphere and easterlies in the summer hemisphere, a result of the strong radiative heating of the summer stratosphere which leads to a reversed temperature gradient compared to  the troposphere. 
Since the stratospheric temperature gradients are larger for the planet around the F-type star the stratospheric zonal wind is stronger for this scenario. 
The distinct temperature increase above the summer pole is missing for the planet around the K-type star, thereby the latitudinal temperature gradient does not reverse at lower pressures and no easterly wind a\-ri\-ses in the summer stratosphere. 
Instead, westerly winds range over a large pressure region throughout the troposphere and stratosphere.  
The zonal wind plays an important role for the upward propagation of planetary waves, which force the meridional circulation. While planetary waves can propagate through westerlies, if the wind speed is not too large, they cannot pass through regions with easterly winds. 
 For the planet around the Sun and the F-type star upward wave propagation is therefore damped in the summer hemisphere by the easterly wind leading to an asymmetric meridional circulation. The absence of a stratospheric easterly wind for the planet around the K-type star allows for the upward propagation of these waves, leading to a more symmetric meridional circulation of the upper atmosphere. A similar pattern as for the planet around the K-type star is found for the planet orbiting the Sun during springtime, where no strong easterly jet is present in the stratosphere. \\ 

It is expected that the change in the temperature structure would be even more pronounced when taking the change in the atmospheric ozone concentrations due to different stellar SEDs into account since 1D modeling studies have shown that ozone concentrations are lower for Earth-like planets around K-type stars, which will lead to even lower stratospheric temperatures, and larger for planets around F-type stars resulting in even higher stratospheric temperatures (Selsis 2000, Segura et al., 2003, Grenfell et al., 2007\nocite{Selsis2000,Segura2003,Grenfell2007}).

Since previous 3D modeling studies of planets around other stars did not consider the heating by ozone, our study is the first to show the impact of the stellar spectral energy distribution upon the temperature structure and the atmospheric dynamics arising from ozone heating for planets around different host stars. While for the planet around the F-type star and the Sun this leads to very different temperature structures to those discussed in the literature, the temperature structure we find for the planet around the K-type star is similar to those found by other studies for atmospheres without ozone \citep[e.g.][]{Shields2013,Shields2014}, showing a decline of temperature with height. In the study presented here the lack of the stratospheric temperature increase leads to a change in the meridional circulation of the upper atmosphere for the planet around the K-type star. This will lead to a change in the transport of trace gas species such as ozone compared to the planet around the Sun, possibly influencing its total column amount, as ozone is known to build up above the winter poles for the Earth. This possible change in the ozone transport, as well as the change in the temperature structure, will have an impact on e.g.~the spectral appearance of such a planet.

\subsection{Hydrological cycle}
\label{sec:hydrocycle}
\noindent
Water vapor plays an important role for the climate of a planet since it is an effective greenhouse gas. It can furthermore undergo phase changes, affecting the atmospheric temperature by the release or storage of latent heat. Water clouds scatter stellar radiation and contribute to the greenhouse effect. Furthermore, liquid water is a prerequisite for life as we know it. A planet with a liquid water reservoir on its surface is expected to also have water in its atmosphere which makes it a very important atmospheric constituent. Additionally the water vapor feedback can intensify climate responses: Higher surface temperatures cause a higher evaporation of water from the surface resulting in an increase of atmospheric water vapor and thereby a larger greenhouse effect and higher surface temperatures. However the overall effect is uncertain as the formation of clouds may alter the response as has been shown by Wolf and Toon (2014), Leconte et al.~(2013b), Yang et al.~(2013, 2014)\nocite{Wolf2014, Leconte2013, Yang2014, Yang2013}. The uncertain 
impact of clouds for other climates has been widely discussed for climate change predictions, see e.g.~\cite{cloudsipcc5}.

\begin{figure}[ht!]
 \centering
{
\includegraphics[angle=-90,width=0.4\textwidth]{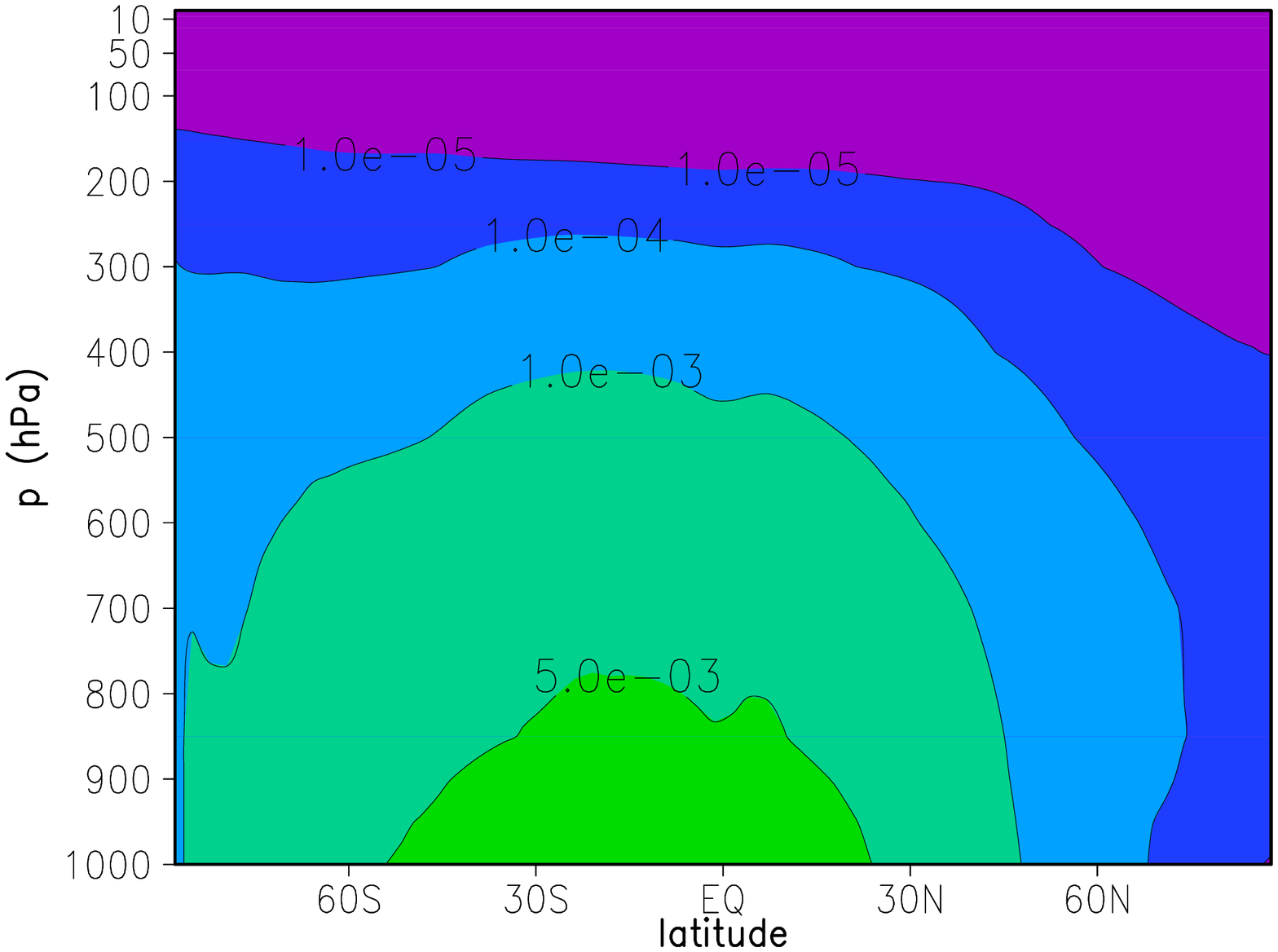}
\includegraphics[angle=-90,width=0.4\textwidth]{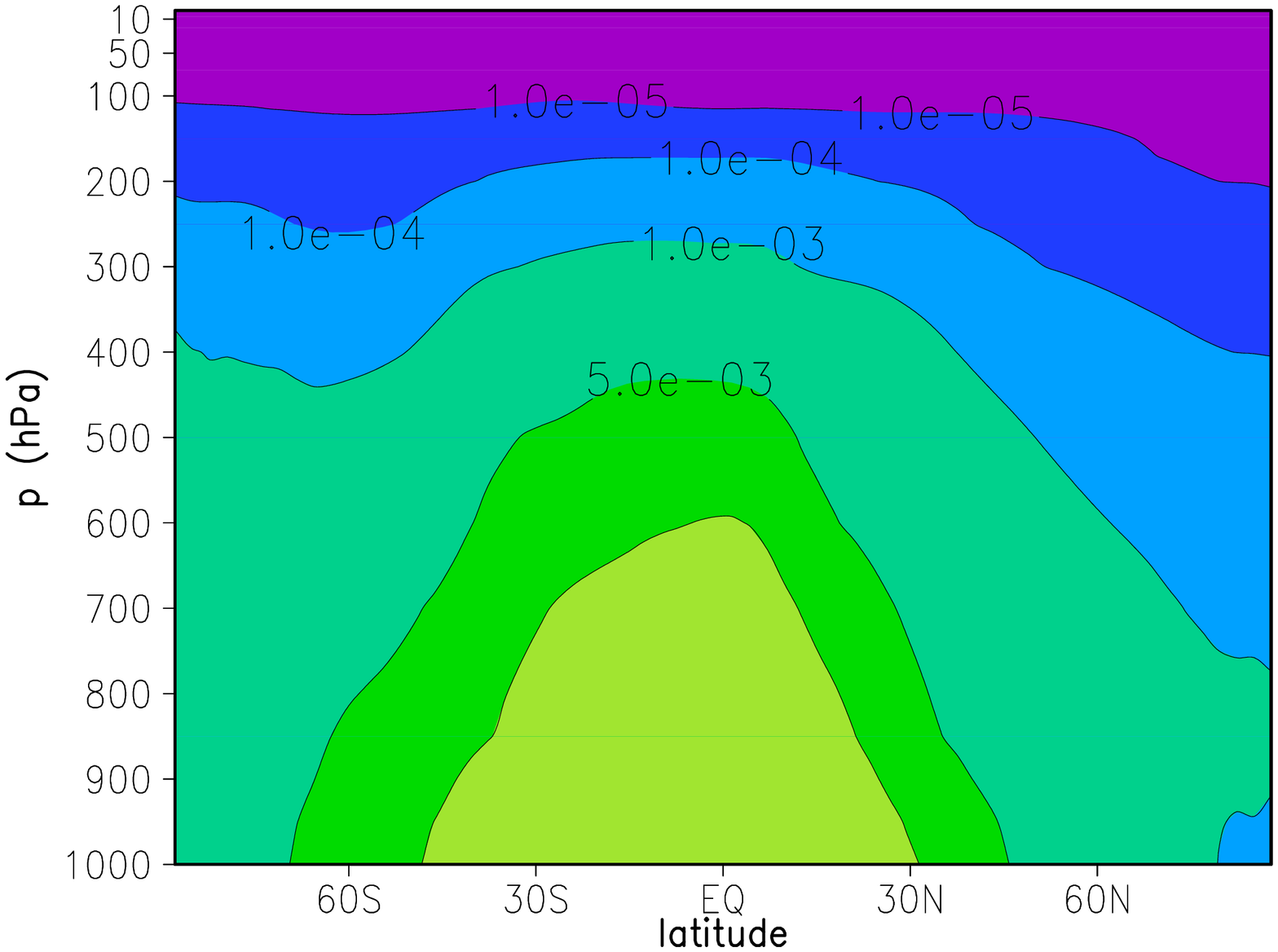}
\includegraphics[angle=-90,width=0.4\textwidth]{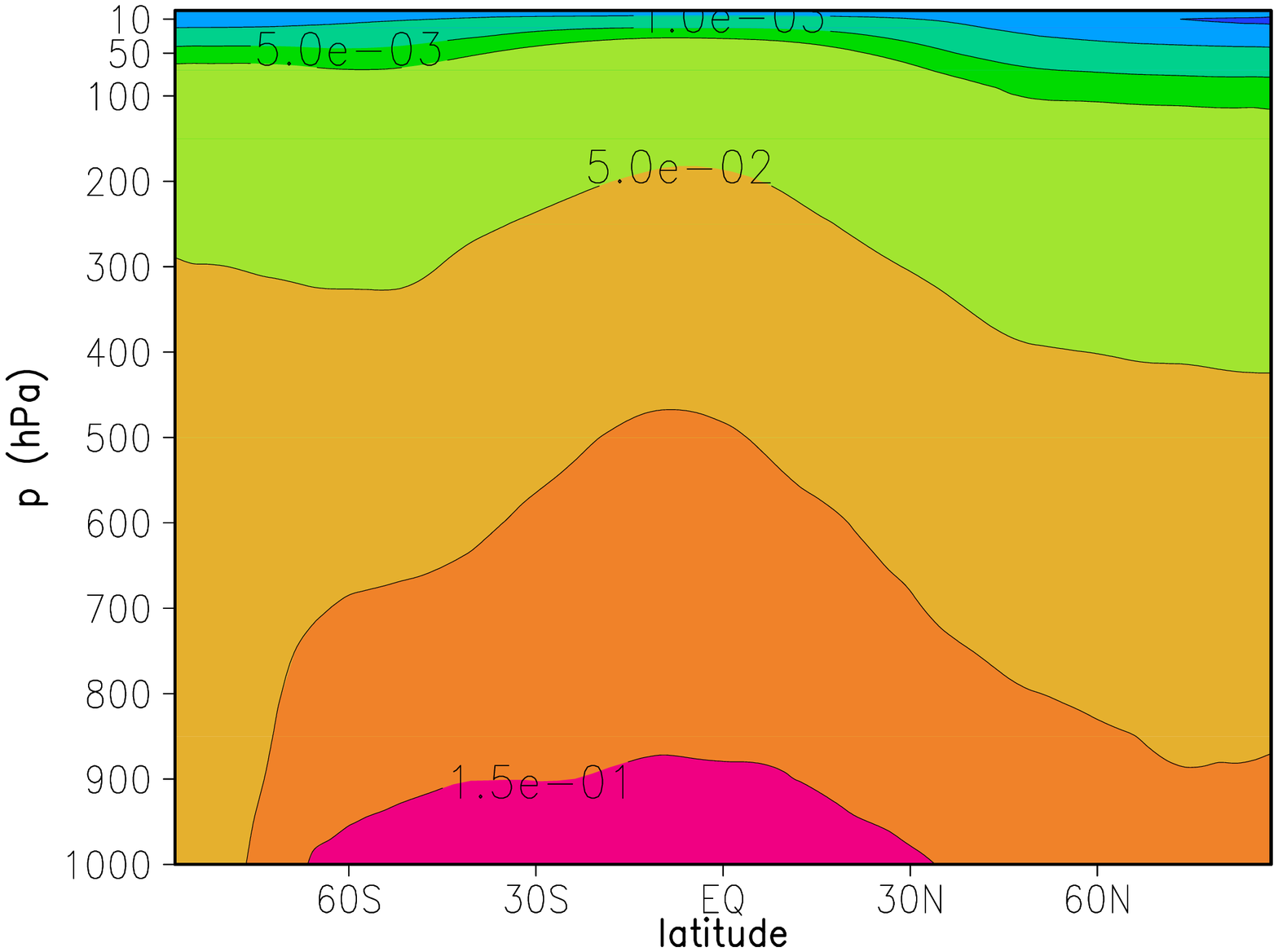}
\includegraphics[angle=0,width=0.5\textwidth]{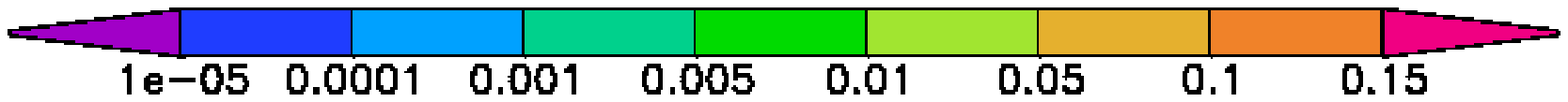}

}
\caption[Zonal mean water vapor distribution for planets around different stars]{Zonal mean distribution of water vapor (vmr) for planets around different stars in NHW. Upper panel: results for planet around the F-type star, middle panel: planet around the Sun, lower panel: planet around the K-type star. Note the linear pressure scale.}
\label{fig:h2o}
\end{figure}

Figure \ref{fig:h2o} shows the zonal mean distribution of the water vapor volume mixing ratio for the three scenarios in northern hemispheric winter. 
For the planet around the Sun the highest volume mixing ratios (vmrs) of around 0.01-0.02 occur for near surface equatorial regions. From low to high latitudes the water vapor mixing ratio decreases steeply due to the decrease in temperature and hence evaporation. The summer hemisphere shows higher mixing ratios as expected.\\
For the planet around the F-type star the water vapor mixing ratio is lower than for the reference scenario. Maximum water vapor vmrs in the summer equatorial regions are lower than 0.01. As for the planet around the Sun, water vapor concentrations steeply decrease towards the poles. Due to the lower water vapor amount in the atmosphere the greenhouse effect becomes less efficient, as indicated by the more negative near surface values of the net thermal radiation in Fig.~\ref{fig:Ftherm}.  \\

\begin{figure}[h]
 \centering 
{\includegraphics[width=0.23\textwidth]{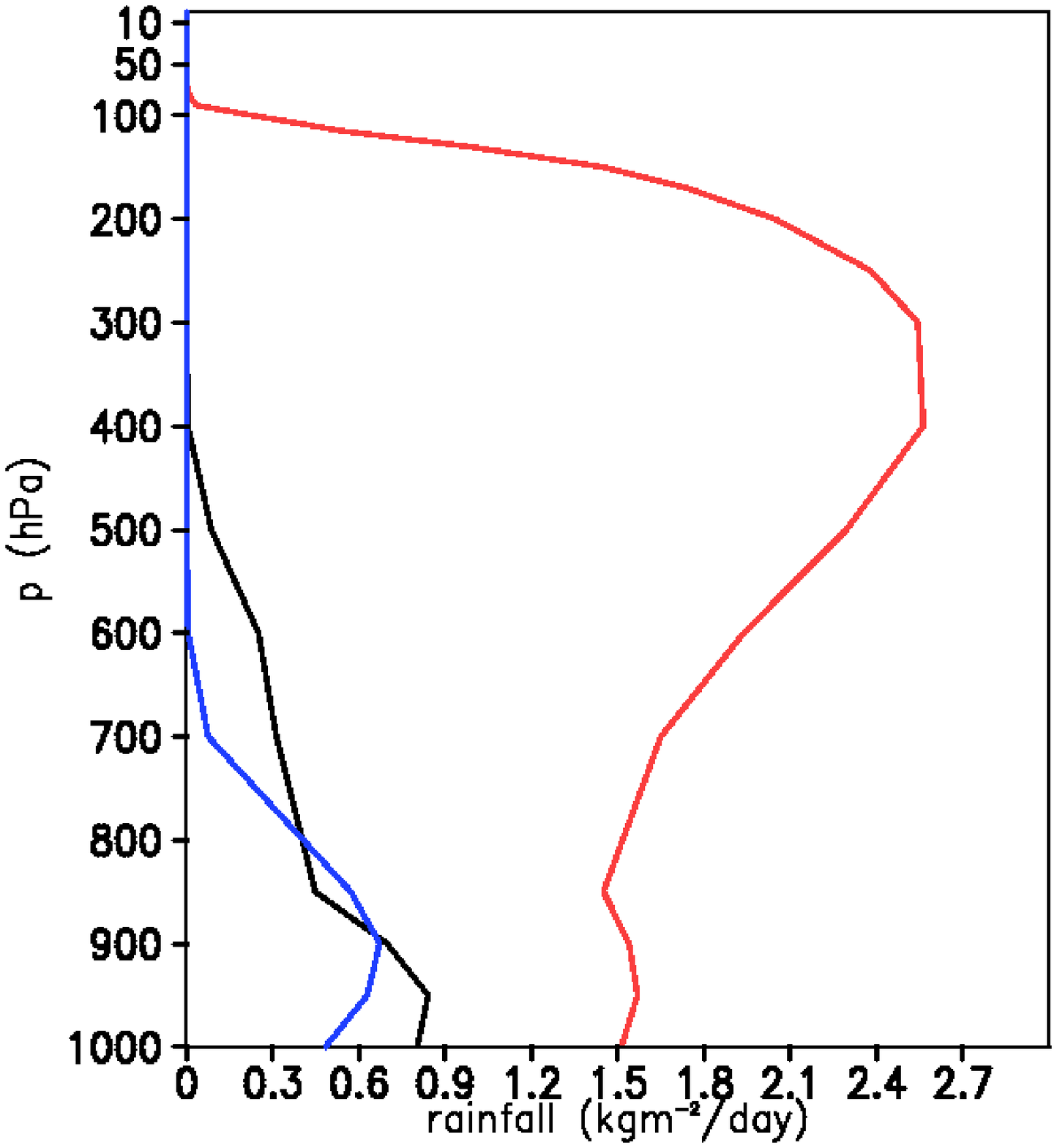}
\includegraphics[width=0.23\textwidth]{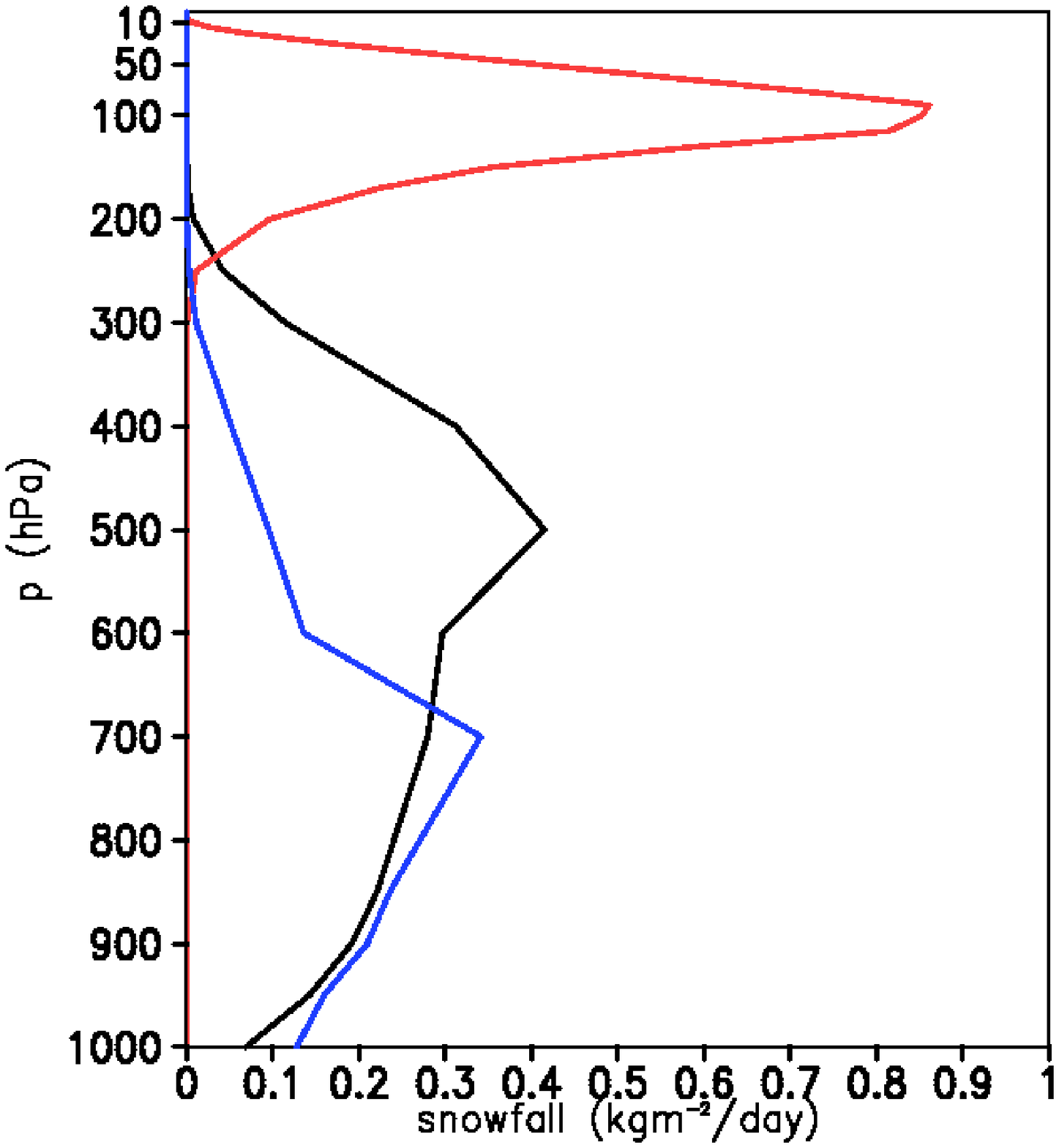}
}
\caption{Global orbital mean rain and snowfall for Earth-like planets around different stars. The left panel shows the rainfall profile and the right panel the vertical snowfall profile. 
The planet around the F-type star is depicted in blue, the planet around the Sun in black, and the planet around the K-type star in red.}
\label{fig:orbgmamprecip}
\end{figure}

For the Earth-like planet around the K-type star much higher water vapor mixing ratios are obtained than for the planet around the Sun, reaching vmrs higher than 0.12 in the equatorial region near the surface. Mixing ratios typical for the surface of the Earth occur at much higher altitudes at pressures of about \unit[100-200]{hPa}. Here the concentrations are nearly constant over all latitudes, which results from a change in the hydrological cycle, which leads to recycling of water in the upper atmosphere. While for the Earth, cloud formation and precipitation lead to efficient replenishment of surface water, Figure \ref{fig:orbgmamprecip} shows that for the planet around the K-type star a large part of the precipitation evaporates or melts before reaching the planetary surface. The high atmospheric temperatures lead to melting of all solid precipitation at pressures larger than~\unit[200]{hPa}. The amount of water recycled by this 
melting and evaporation processes in the upper atmosphere constitutes a reservoir larger than the precipitation reaching the surface for the planet around the Sun. Nevertheless rainfall at the surface is intensified on the planet around the K-type star compared to the reference scenario. 
Thus, these high water vapor concentrations are the result of a change in the water vapor feedback cycle and the hydrological cycle. \\
The high water vapor mixing ratios for the planet around the K-type star are also a result of the assumed land-sea mask, as about 70\% of the planet assumed is covered with water. For a planetary scenario with a much smaller water reservoir, e.g.~ a planet without any ocean, the impact of the hydrological cycle upon surface temperatures would be much smaller as discussed e.g.~by \cite{Abe2011}, since the greenhouse effect of water vapor would be much weaker in such a scenario. For a planetary scenario with a larger ocean reservoir, we expect a similar response of the hydrological cycle as the amount of water vapor in the atmosphere is related to the surface temperature via phase equilibrium. Due to a lower surface albedo for an ocean planet, surface temperatures would probably be higher. 

A comparison of the water vapor concentrations for the planet around the K-type star with model results of \cite{Kasting1993hz}, who investigated the water loss limit as an inner boundary of the habitable zone shows that the stratospheric water vapor vmr (ca.~\unit[2$\times10^{-4}$]{}) is still below their critical value of \unit[3$\times10^{-3}$]{}. Thus, according to these results a water reservoir of one Earth ocean should be stable over \unit[4.5]{Gyrs} when considering diffusion limited escape of hydrogen to space for this scenario. \cite{Kasting1983} however calculated atmospheric concentrations in the upper atmosphere of a Venus-like terrestrial planet, starting with stratospheric water vapor concentrations similar to those obtained here and found that an increase in the solar EUV flux by a factor of ten may result in strong hydrogen escape. Therefore, it is possible that even though the water vapor concentrations are below the critical value calculated by 
\cite{Kasting1993hz} a different stellar 
spectral energy distribution, especially with increased high-energy radiation may still lead to a water loss for such a scenario. 

In addition to water vapor, also clouds can have a strong impact on exoplanetary climate as shown e.g.~by \cite{Kitzmann2010}. We cannot judge whether the formation of clouds in Earth-like atmospheres of rocky extrasolar planets is captured by the cloud scheme included in our 3D model, since even for Earth climate predictions the impact of clouds varies over a wide range \citep{cloudsipcc5}. A first step towards the understanding of the cloud feedback for rocky extrasolar planets could be a model comparison for such scenarios. For the planet around the F-type star we find an increase in cloud cover and a decrease in cloud cover for the planet around the K-type star compared to the planet around the Sun, see also table \ref{tab:gm_orb}. We find the same behavior for the water ice column. For the cloud water column however we find an increase for the planet around the K-type star and a decrease for the planet around the F-type star, see table \ref{tab:gm_orb}. The cloud water column is about twice as high for the planet around the K-type star compared to the planet around the Sun. The greenhouse effect caused by clouds (Cloud GHE in table \ref{tab:gm_orb}) is calculated from the difference in surface temperature and the effective temperature from the longwave radiation of the planets for cloudy and clear sky conditions. Our results suggest the cloud GHE is larger for the planet around the K and the F-type star than for the planet around the Sun. This indicates that the influence of clouds critically depends on cloud properties and atmospheric temperatures, and cannot be easily estimated without sophisticated cloud modeling. The differences in cloud impact of different Earth climate models upon Earth climate predictions, but also studies of exoplanet scenarios, such as e.g.~\cite{Wolf2014, Leconte2013},  show that much more work is needed to understand the behavior and influence of clouds for the Earth and for exoplanetary climates. For the planet around the K-type star, the cloud GHE constitutes only about 15\% of the total GHE, while for the planet around the F-type star it is about 43\% and 22\% for the planet around the Sun. Hence, for the warm scenario the cloud GHE seems to be of minor importance. 

\subsection{Surface conditions}
\label{sec:surf}
\begin{figure}[ht!]
 \centering
{\includegraphics[angle=-90,width=0.4\textwidth]{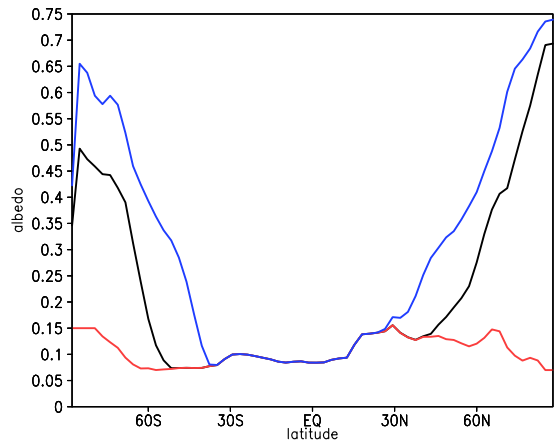}
}
\caption[Surface albedo for planets around different stars]{Zonal orbital mean surface albedo for planets around different stars. 
The planet around the F-type star is depicted in blue, the planet around the Sun in black, and the planet around the K-type star in red.}
\label{fig:surf}
\end{figure}

\noindent
The change in temperature is linked to a change in the surface albedo via the albedo feedback. Higher temperatures lead to melting of sea ice whereas lower temperatures will lead to its buildup. Figure \ref{fig:surf} shows the zonal orbital mean surface albedo for the three scenarios studied.  
At low and mid latitudes the albedo is the same for all scenarios. At mid to high latitudes the albedo differs depending on snow and ice coverage.\\
For the planet around the K-type star no sea ice or snow is present, leading to a uniformly low surface albedo.\\
 The highest albedos occur for the planet around the F-type star at the north pole, which is a result of strong buildup of sea ice. In the southern hemisphere highest albedos occur also for the planet around the F-type star at latitudes where sea ice is present.

For the planet around the F-type star the ice-albedo feedback causes an increase in surface albedo due to buildup of sea ice down to latitudes of about 40$^{\circ}$. Despite the fact that water vapor concentrations are low, i.e.~the greenhouse effect of water is smaller than for the planet around the Sun, the increase in surface albedo is not large enough to cause a runaway glaciation. Note however, that depending on the stellar SED, the effect of the surface albedo may differ, since especially the albedo of snow is strongly wavelength-dependent, as e.g.~shown by \cite{Joshi2012}. In our study we do not account for the wavelength dependence of the surface albedo. Including it may increase the ice-albedo feedback for the planet around the F-type star as shown by \cite{Shields2013}.

\subsection{Sensitivity tests}
\label{sec:sensitivity}
\begin{figure}[ht!]
 \centering
{\includegraphics[width=0.4\textwidth]{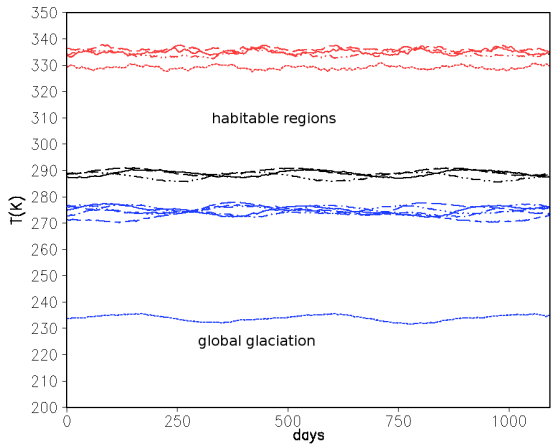}
}
\caption{Influence of different q-fluxes and orbital periods on near-surface temperature for planets around different stars. Results for the planet around the F-type star in blue, for solar radiation in black, and for the planet around the K-type star in red. q-flux corrections: q1 dot-dot dashed, q2 solid, q3 long dashed, q4 dot-dashed for the F3Dq4 scenario with a \unit[450]{days} orbit, q4 long dashed-short dashed for the F3D scenario with a \unit[868]{days} orbit, q5 dotted.}
\label{fig:qflux}
\end{figure}

\noindent
For an extrasolar planet, besides many other parameters, also the oceanic heat flux term is a-priori unknown. It was shown, by e.g.~\cite{Yang2013}, \cite{Hu2013} and \cite{Cullum2014}, that the oceanic heat flux may strongly influence atmospheric GCM results and surface temperatures. Therefore, we test the sensitivity of our model results to a change in the q-flux term. Additionally, we vary the orbital period to estimate its influence on the model results, since different lengths in season may lead to different results due to the large thermal inertia of the ocean.\\ 
The scenarios and their resulting global orbital mean temperature are summarized in table \ref{tab:sensitivity}. Overall, the difference in temperature due to different orbital periods is small, about \unit[1]{K} for the planet around the F-type star (comparing scenarios F3D and F3Dq4, which both use the q4 q-flux but different orbital periods).  However, the seasonality changes, hence for longer orbital periods, longer temperature maxima and minima are obtained, see Fig.~\ref{fig:qflux}. For the planet around the F-type star and the Sun the seasonality is more pronounced than for the planet around the K-type star. While for the planet around the K-type star the atmosphere is dominated by the large amount of water vapor, for the planet around the F-type star and the Sun the build-up and melting of sea ice are important factors for the seasonal temperature variability. Note that for the planet around the F-type star we conducted most of the sensitivity tests with an orbital period of \unit[450]{days}, which was computationally less expensive than using the physically correct orbital period of \unit[868]{days} used for the results in the previous sections .

\begin{table*}
\caption[Sensitivity]{Sensitivity test scenarios and global orbital mean near-surface temperatures.}
\begin{center}
 \begin{tabular}{lllll}
Scenario&Star (stellar type)  & orbital period (d) & q-flux term (variability)& T$_{2m}$ (K)  \\
\hline\hline
F3D & $\sigma$ Boo (F2V)&868.64&q4 (am)& 273.6     \\  %mgfs1mq
F3Dq0 & $\sigma$ Boo (F2V)&450.0&q5 (=0)& 234.0     \\%fst0q
F3Dq4 & $\sigma$ Boo (F2V)&450.0&q4 (am)& 274.7     \\%fstamq
F3Dq3 & $\sigma$ Boo (F2V)&450.0&q3 (mm)& 275.5     \\%forbmq
F3Dq2 & $\sigma$ Boo (F2V)&450.0&q2 & 275.2     \\
F3Dq1 & $\sigma$ Boo (F2V)&365.25&q1 &275.0     \\\hline

Sun3D &  Sun (G2V)&365.25&q2 &  288.6   \\ %Sunorb
%Sun3Dq4 &  Sun (G2V)&364.25&q4&  288.6   \\
Sun3Dq3 &  Sun (G2V)&365.25&q3 (mm)&  289.7   \\ %mgsunorbmq
Sun3Dq1 &  Sun (G2V)&365.25&q1 &  287.7  \\ %mgsunmlo
\hline

K3D& $\epsilon$ Eri (K2V)& 184.00&q2 &334.9      \\
K3Dq0& $\epsilon$ Eri (K2V)& 184.00&q5 (=0)&329.2     \\%klo0q
K3Dq2& $\epsilon$ Eri (K2V)& 184.00&q3 (mm)& 336.2     \\ %korbmq
K3Dq1& $\epsilon$ Eri (K2V)& 365.25&q1 &334.5    \\ %klomloalb

\hline
\end{tabular}
\label{tab:sensitivity}
\end{center}
\end{table*}

The influence of varying the oceanic heat flux (q-flux), used in the mixed layer ocean model to calculate the sea surface temperatures and sea ice, is a little larger, leading to a temperature difference of up to \unit[2]{K} for moderate changes in the q-flux term. This term partly represents the oceanic circulation and heat transport.  We applied five different q-flux terms q1-q5, as described in sec.~\ref{sec:scenarios}. Figure \ref{fig:qflux} suggests that moderately varying the q-flux term (shown are q2-q4) does not change the global temperature response. A seasonal variation of the temperature can be identified for the planet around the Sun and the F-type star. For the planet around the K-type star the the variation of the temperature is less pronounced. A larger variability at smaller timescales is visible which is less periodic. This is caused by the large amount of water compounds in the atmosphere which are highly variable on timescales smaller than the seasons.\\

While on the global scale the near surface temperatures deviate by up to \unit[10]{K} for different q-fluxes and orbital periods (except for the planet around the F-type star with a q-flux of 0), on the regional scale the temperature differences are larger. For the planet around the F-type star we find a temperature difference of about \unit[30-50]{K} at the summer pole and of about \unit[20]{K} for the winter pole due to the difference in orbital period, with higher temperatures in summer and lower temperatures in winter for the longer orbital period (not shown). For the planet around the Sun the different q-fluxes q1-q3 lead to a small change in the global annual mean near surface temperatures from an exoplanet perspective. For Earth climate calculations these differences of about \unit[2]{K} are however large. On the local scale we find e.g.~larger near surface temperatures (of about \unit[5]{K}) above the equatorial ocean and the southern pole for the q-flux q3. The zonal seasonal mean temperature structures are not influenced strongly by the change of the orbital periods or q-fluxes either, from an exoplanet perspective. Compared with the scenarios with an orbital period of 365 days and the q-flux q1 we find a very similar temperature structure, hydrological cycle, and a similar dynamical behavior, and surface response compared to those presented in the previous sections. It was not anticipated that the change in the orbital period and the oceanic heat flux would have such a small effect on the mean 3D model results.\\

Only the assumption of an extreme and unrealistic q-flux of zero (q5) leads to a strong change in surface conditions for the planet around the F-type star, which undergoes global glaciation. Hence, for this scenario, an oceanic heat flux is required to prevent the planet from freezing completely. For the planet around the K-type star the orbital mean temperature only decreases by about \unit[5]{K} for the q5 q-flux, since the greenhouse effect of the atmosphere dominates the heat budget. \\
In \cite{Shields2013} a similar scenario has been studied: an Earth-like planet around an F-type star with an oceanic heat flux of zero. They discuss that their planet is close to global glaciation for a total amount of incoming stellar radiation as Earth receives from the Sun. Hence, although they include a wavelength-dependent ice albedo in their study, which should enhance the response by increased scattering of stellar light, in their model calculations the planet does not undergo global glaciation. Their model results differ from ours because they disregard ozone in their calculations and assume a surface completely covered with water which lowers the background surface albedo. Furthermore, the albedo they assume in the visible wavelength regime (0.8 cold dry snow) is about equal to the one we assume in our model calculations for the entire shortwave regime (0.8 for snow on ice). Their albedo in the NIR regime is however lower (0.68 for snow on ice).  Additionally, from their paper it is unclear whether they also use different amounts of greenhouse gases, such as CO$_2$ and CH$_4$, which may also yield higher surface temperatures.

\subsection{Comparison of the 3D to the 1D model results}
\label{sec:1d3d}
\noindent
To evaluate the habitability of rocky extrasolar planets 1D atmospheric models are often utilized since the 1D assumption keeps the number of boundary conditions small. Furthermore, in future, averaged quantities of these distant worlds will be the first to be retrieved. However, it has been suggested that for rocky extrasolar planets close to the inner edge of the habitable zone 3D phenomena may have to be taken into account for certain scenarios of e.g.~tidally-locked extrasolar planets or for planets with small water reservoirs (Abe et al., 2011, Leconte et al., 2013b, Yang et al., 2014\nocite{Abe2011,Leconte2013, Yang2014}).\\ 

\begin{figure}[ht!]
 \centering
{\includegraphics[width=0.45\textwidth]{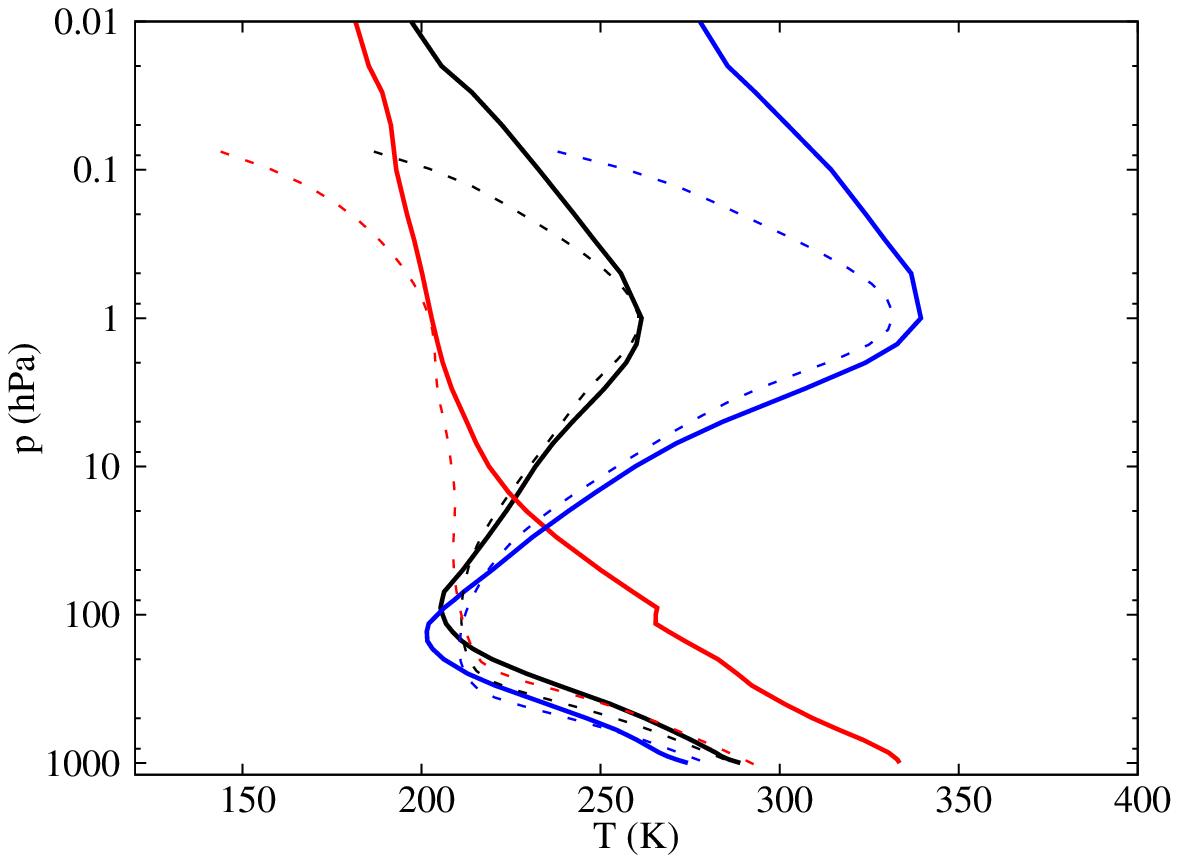}
\includegraphics[width=0.45\textwidth]{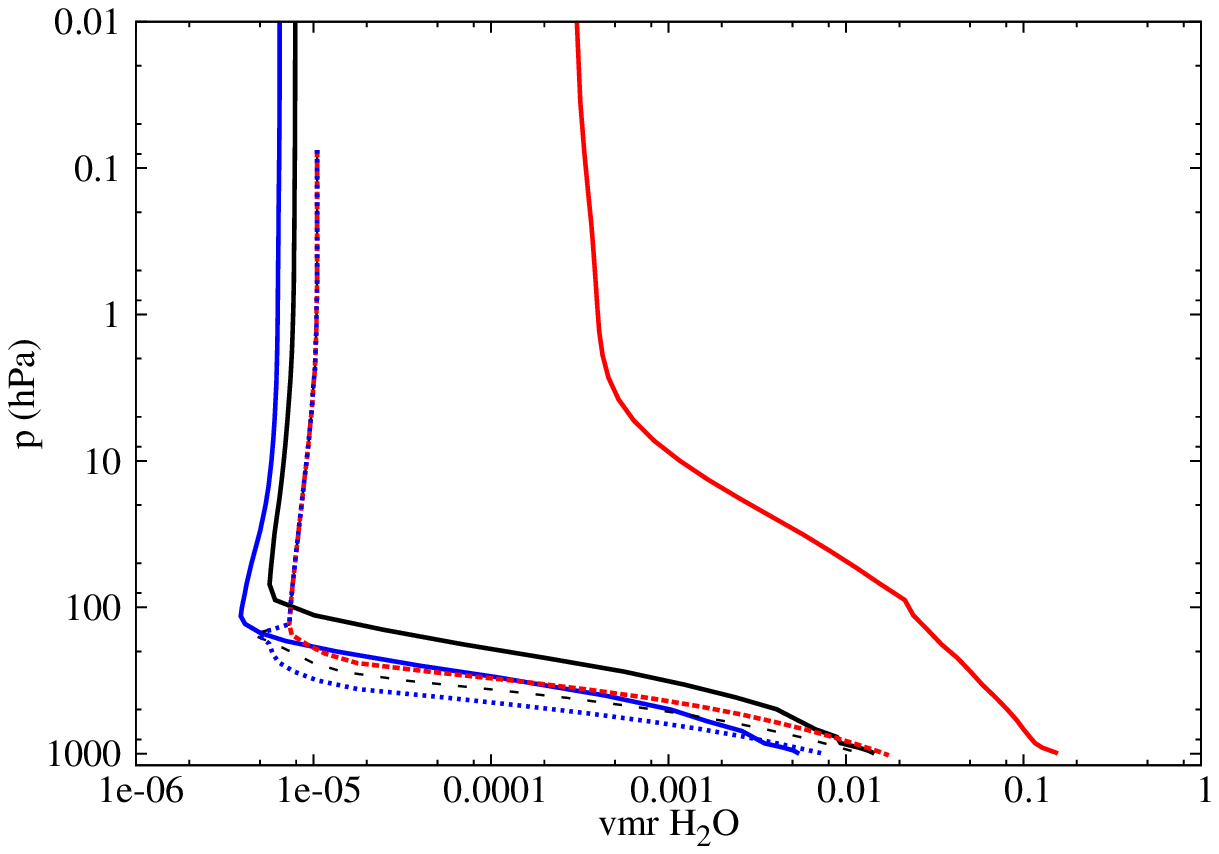}}
\caption{Mean temperature (top) and water vapor (bottom) profiles for Earth-like planets around different types of stars. Global orbital mean results from the 3D model in solid, profiles from the 1D model in dashed lines. Results the planet around the F-type star in blue, for solar radiation in black, and for the planet around the K-type star in red.}
\label{fig:1dth2o}
\end{figure}

Here we test whether 3D phenomena need to be taken into account for habitable Earth-like extrasolar planets which are not tidally locked by comparing the global orbital mean results of the 3D model with those of a steady-state radiative-convective model (see section \ref{sec:1Dmodel}). The upper panel in Fig.~\ref{fig:1dth2o} shows the temperature-pressure profiles for the three scenarios as computed with the 3D and the 1D models.\\

For the planets around the Sun and around the F-type star, the temperature profiles compare well for pressures between \unit[1000]{hPa} and $\approx$\unit[1]{hPa}. Differences at larger height may be explained by the different model regimes. In particular, the 3D model extends to \unit[0.01]{hPa} while the 1D model only ranges to \unit[0.066]{hPa}. Also in the 1D model, e.g.~the absorption of radiation at \unit[121.5]{nm} (Lyman $\alpha$) by molecular oxygen is not considered leading to lower temperatures in the upper model atmosphere.
For the planet around the K-type star a larger difference between the global orbital mean profile of the 3D model calculations and the 1D model can be found. This mainly results from the large difference in the calculated water vapor profiles, which are shown in the lower panel of Fig.~\ref{fig:1dth2o}. In the 1D model the water vapor profile is determined from an assumed vertical profile of relative humidity. For the 1D model calculations shown in Fig.~\ref{fig:1dth2o} (F1D, Sun1D, K1D) a relative humidity profile as measured for present Earth \citep{MW1967} is assumed for the calculation of the water vapor profile, which starts with a relative humidity of \unit[77]{\%} at the surface and then decreases throughout the troposphere. This results in the steep decline with height of water vapor in the atmospheres as calculated by the 1D model. However, in the 3D model calculations, as shown in section \ref{sec:hydrocycle}, the distribution of water vapor in the atmosphere differs due to a change in the hydrological cycle. 
Therefore, the water vapor profile and thus also the temperature profiles as well as the global mean surface values (Tab.~\ref{tab:1D3D_surf}) calculated by the 1D and the 3D model disagree for the planet around the K-type star.\\

\begin{table}[ht!]
\caption[1D--3D temperature and water vapor for planets around different stars]{1D and 3D global mean temperature (T) and water vapor vmr (H$_2$O) at \unit[1000]{hPa} for the Earth-like planets around different stars as calculated in this study, \textit{Star}1D/3D, and in Stracke (2012) with different relative humidity profiles: RHMW from \cite{MW1967} RH100 with a relative humidity of 100\%. }
\begin{center}
 \begin{tabular}{llll}
  Scenario& T (K) &  H$_2$O (\%)\\
\hline\hline
F3D &274.3 &0.6 \\
F1D & 280.1 &0.8 \\\hline
Sun3D  &  289.0 &1.4 \\
Sun1D   &  287.7 & 1.3\\
\hline
K3D    &  333.4 &15.6 \\
K1D    & 292.9 &1.7 \\
K1DRHMW &292.0  & 1.7\\
K1DRH100 & 329.0&14.1\\ 
\hline

 \end{tabular}
\label{tab:1D3D_surf}
\end{center}
\end{table}

\begin{figure}[ht!]
 \centering
{\includegraphics[width=0.5\textwidth]{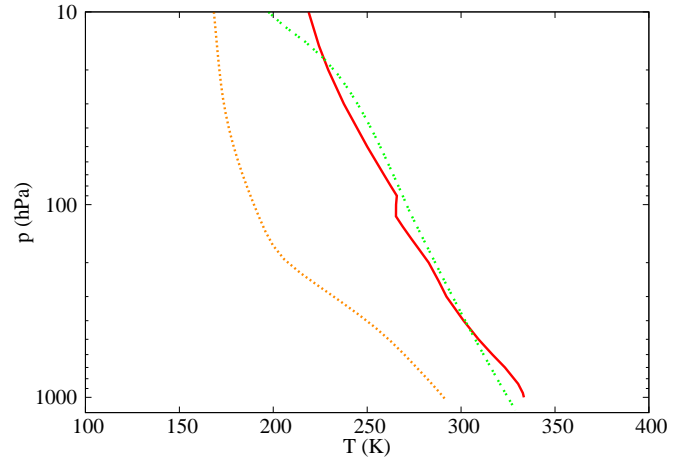}}
\caption{Comparison of the global orbital mean tropospheric temperature profile of the 3D model results for the planet around the K-type star (red, solid) with 1D model results \citep{Stracke2012} for the corresponding scenario with different RH profiles: orange, dotted with a relative humidity profile from \cite{MW1967}, green, short-dashed with a relative humidity of \unit[100]{\%}}
\label{fig:kstarT_1D3D_rh}
\end{figure}

To test whether the 3D global orbital mean profiles can nevertheless be approximated by the 1D model, we additionally compared the 3D model results for the K-type star with 1D model profiles calculated with a relative humidity of \unit[100]{\%} by \cite{Stracke2012} (K1DRH100). This study utilized the same atmospheric model however with e.g.~an improved thermal radiative transfer scheme (applicable to larger temperature and pressure range, \cite{vonParis2010}), and applied it to an Earth-like planet around different types of stars however neglecting the influence of oxygen, ozone, methane, and nitrous oxide. For these model calculations (K1DRHMW, K1DRH100) a surface al\-be\-do of 0.22 was assumed. As can be inferred from Fig.~\ref{fig:kstarT_1D3D_rh}, for the assumption of a fully saturated atmosphere (K1DRH100) the tropospheric profiles of the 3D model results for the planet around the K-type star and 1D model calculation from 
\cite{Stracke2012} agree well. Surface temperatures and water vapor volume mixing ratios also compare better for this assumption, see table \ref{tab:1D3D_surf}.\\

This implies that the 1D model can indeed approximate the global orbital mean climate states calculated with the 3D model, when assuming an appropriate relative humidity profile. Note, that assuming a relative humidity of \unit[100]{\%} in the 1D model increases the surface temperature of the planet around the Sun to \unit[305]{K} (for an atmosphere with N$_2$, CO$_2$, and H$_2$O). Thus the agreement of the 3D and 1D model results strongly depends on the choice of the relative humidity profile used for the water vapor profile calculation. Which profile to choose is however a-priori unknown. 

Note that the realization of a global mean relative humidity of \unit[100]{\%} seems unlikely as atmospheric dynamics will lead to drying of air during ascent \citep{Leconte2013b}, by condensation and cloud formation for most scenarios.
In the 3D model calculations presented in sec.~\ref{sec:temperatures}--\ref{sec:surf} we find a global mean surface relative humidity of only 74\% for the planet around the K-type star (71\% for the planet around the Sun and 69\% for the planet around the F-type star), lower than the relative humidity of 100\% assumed in the 1D model to find a good agreement between the results. However, the decrease in relative humidity with height is weaker for the planet around the K-type star than for the planet around the Sun and the F-type star in the 3D model results. For these cooler scenarios the decrease is similar to the RH profile  by \cite{MW1967} assumed in the 1D model. Including the global mean relative humidity profile from the 3D model calculations in the 1D model results in an increased surface temperature which reaches \unit[315]{K}. We find a similar good agreement between the 1D and 3D model results as with a constant RH of 100\% in the 1D model, when assuming a fixed water vapor profile corresponding to the 3D global mean water vapor profile and a global mean surface albedo of 0.1 in the 1D model calculations for the planet around the K-type star, yielding a surface temperature of \unit[328]{K}.  
\\

From our 3D model calculations which include the change in the hydrological cycle, the water vapor feedback, and the ice-albedo feedback, we deduce a  maximum difference in the global orbital mean near-surface temperature of \unit[61]{K}, for the planet around the F-type star and the planet around the K-type star (see table \ref{tab:1D3D_surf}) when including an Earth-like oceanic heat transport. Including also the q-flux sensitivity test scenarios (see section \ref{sec:sensitivity}) the maximum difference is \unit[102]{K} for the same amount of total energy incident at top of the atmosphere. The largest difference in near surface temperatures obtained by the 1D model calculations in this work is \unit[49]{K}, for the planet around the F-type star with a relative humidity profile following \cite{MW1967} and the planet around the K-type star with a relative humidity of 100\%. The maximum temperature difference obtained by the 1D model calculations is smaller because a change in the surface albedo is not captured by the 1D model calculations. Note that the 3D model results depend on complex parameterizations which are adjusted to reproduce present and past Earth climates. Therefore, the results may differ in absolute numbers for varying the parameter sets and model codes. Our results should hence be considered to be of more of a qualitative than of quantitative nature.

In 1D model studies, similar to that presented here, the surface albedo is often tuned to fit the mean temperature of the Earth (\unit[288]{K}), and therefore not only includes the real surface albedo, but additionally mimics the reflectivity of clouds within the atmosphere. The surface albedo in the 3D model is calculated and differs for the scenarios investigated here as discussed in section \ref{sec:surf}. As long as the planets stay habitable the difference in surface albedo is however relatively small ($\Delta$albedo $\sim$ 0.1). When including the sensitivity scenario where the planet around the F-type star undergoes global glaciation the difference in albedo is large ($\Delta$albedo $\sim$ 0.5). The impact of clouds on the planetary albedo can be very diverse \citep[see e.g.][]{Kitzmann2010}, and therefore difficult to capture. 

Which surface albedo  is realized for a certain planetary scenario needs to be calculated interactively by a coupled 3D atmosphere-ocean general circulation model (AOGCM), since the resulting surface albedo strongly depends on the oceanic heat transport as illustrated in section \ref{sec:sensitivity}. Therefore atmospheric GCMs coupled to a mixed layer ocean, as utilized in this work, may only give a first estimate of the surface albedo for a certain planetary scenario. It would preferably be calculated with coupled atmosphere-ocean models. Such calculations will however introduce an even larger amount of parameters which are unknown for extrasolar planets, and are more expensive in terms of computing time. Their impact should nevertheless be explored. 
It should be noted that the planetary surface albedo is important to deduce the climate of planets with relatively thin 
atmospheres, as for these planets it has large impact on the energy budget. For planets with thicker atmospheres e.g. at the inner edge 
of the habitable zone, where one may expect water dominated atmospheres \citep[e.g.][]{Stracke2012} or at the outer edge of the habitable zone where one may expect thick CO$_2$-do\-mi\-nat\-ed atmospheres its impact  is much smaller because it is masked by the thick atmospheres \citep{vonParis2013ice,Shields2013} .

\section{Summary and conclusion}
\label{sec:sum}
\noindent
The results of 3D GCM climate modeling of Earth-like planets around F, G, and K-type stars have been discussed in this paper. It has been shown that different stellar spectral energy distributions may lead to very different climate states of an Earth-like planet for the same total amount of stellar energy incident at the top of the atmosphere. This results from the wavelength-dependent absorption and scattering properties of the atmosphere, and an amplification of the climate response by positive climate feedback cycles, namely the water vapor and the ice-albedo feedback.\\
For the planet around the F-type star we find an enhanced heating of the stratosphere by ozone due to the change in the stellar energy distribution compared to the planet around the Sun. For the planet around the K-type star we find a strong change in the vertical temperature structure with no temperature increase in the stratosphere, caused by decreased heating by ozone absorption, and a change in the hydrological cycle due to the high tropospheric temperatures.
For nearly all scenarios studied, the Earth-like planets result in habitable surface conditions. Only when neglecting the (parametrized) oceanic heat transport for the planet around the F-type star, which is rather unphysical, we find uninhabitable surface conditions. For this scenario the planet undergoes global glaciation. A moderate change in the oceanic heat flux term and in the orbital period do not show a strong impact on the mean planetary surface climate.\\
Comparing the global orbital mean of the 3D model results to those of a cloud-free 1D radiative-con\-vec\-tive column model showed that the temperature response may be approximated by the 1D model. The agreement of the results, however, crucially depends on the choice of the relative humidity profile utilized to calculate the water vapor profile.

\section*{Acknowledgments}
\noindent We thank the anonymous referees for their helpful comments on the manuscript. This work has been partly supported by the For\-schungs\-allianz \textit{Planetary Evolution and Life} and the Postdoc Program "Atmospheric dynamics and Photochemistry of Super Earth planets" of the Helmholtz Gemeinschaft (HGF). This study has partly received financial support from the French State in the frame of the "Investments for the future" Programme IdEx Bordeaux, reference ANR-10-IDEX-03-02. The  3D model calculations have been performed on the North-German Supercomputing Alliance (HLRN) parallel supercomputing system.
We would like to thank S.~Dietm\"uller and M.~Ponater for providing and advising us with the mixed layer ocean. 

\bibliographystyle{elsarticle-harv}
\bibliography{KGF_paper_v2}

\end{document}